\documentclass[pre,showpacs,preprint,amsmath,amssymb,aps]{revtex4-1}

\usepackage[dvipdfmx]{graphicx}
\usepackage{dcolumn}
\usepackage{bm}
\usepackage{color}
\usepackage{amsmath}

\begin{document}
 
\title{
Self-Assembly Formed by Spherical Patchy Particles
with Long-Range Attraction 
}

\author{
Masahide Sato}
\affiliation{
Information Media Center, Kanazawa University,
Kakuma--machi, Kanazawa 920-1192, Japan
}%

\date{\today}

\begin{abstract}
We report on self-assemblies formed from  spherical patchy particles
interacting by  a long-range attraction through a patch region 
in a two-dimensional system.
We performed Monte Carlo simulations to find stable structures in 
a system with constant number of particles under constant temperature and constant pressure (NPT system),
in which particles interact  via the Kern--Frenkel potential. 
For long-range attractive potentials,
we describe how these stable structures and their formation 
depend on the coverage of the patch.
Under high pressure,
when the coverage is small, triangular lattices are formed as reported in previous papers.
From our simulations, we find when the pressure is low
short chain-like structures, in which the distance between particles is long,
and square clusters,  which are not formed with a short-range attractive potential, 
are formed.
When the coverage of the patch region is large, 
square clusters are formed
since the interaction between particles is stronger than that for with small coverage.
When the coverage ratio is larger than 0.5,
the direction of the patch is perpendicular to the plane in which the particles are placed.
\end{abstract}


\maketitle

\section{Introduction}\label{sec:introduction}

Aggregations and self-assembles formed with  anisotropic particles show characteristic structures
and properties which are not displayed by isotropic particles.
Hence,  anisotropic particles are promising candidates as  building blocks
of functional materials~\cite{Zhan-Nanolett4_2004_1407,Wang_Nature491_2012_51,
Maldovan_natMater3_2004593,Roh_natMater4_2015_759,Mao-G_natMater12_2013_217}.
Recently, many groups designed the anisotropy of colloidal particles by controlling 
the shape of particles~\cite{Glotzer_natMater6-557_2007,Sacanna_nature464_2010_575,Matthew-natMater9_2010_913,
Kraft-J.Phy.Chem.B115_2011_7175,Kraft-pnas109_2012_10787,%
Avvisat-jcp142_2015_084905,Geuchies-natMater15_2016_1,Wolters-Langmuir33_2017_3270,Kang}
 and changing the properties of  an area of the particle's surface. 
When  surface  properties of a particle  are partially changed, the particles are termed patchy.

There have been many studies on how to synthesize patchy particles  and what kinds  of self-assembles are produced 
with them~\cite{Miller-PRE_2009_021404,Chen-Nature469_2011_381,Chan-Science331_2011_199,%
Chen-Langmuir28_2012_13555,Romano-jpcm24_2012_064113,Vissers-psds-jcp138_2013_164505,%
Preisler-vsms-jpcB117_2013_9540,Vissers-JCP140_2014_144902,Preisler-vmsf_solfmatter10_2014_5121,%
Shin-Schweizer_softmatter10_2014_262,Iwashita-k_softmatter10_2014_7170,Preisler-vss-jcp145_2016_064513,%
Iwashita_Scientific_reprot6_2016_27599,Gong_nature550_2017_234,Patra-PRE96_2017_022601}. 
For example,
Vissers and co-workers performed Monte Carlo simulations and studied 
crystals formed from Janus particles~\cite{Vissers-psds-jcp138_2013_164505}, 
namely, spherical particles for which the two halves of the surface have different chemical compositions.
By controlling the pressure and the strength of the attractive interaction between the patch areas of the particles, 
a phase diagram for the structures formed from Janus particles was then developed.
The formation of tubes and polymerization of one-patch particles,
which are the particles having just one patch region, 
has also been studied~\cite{Preisler-vsms-jpcB117_2013_9540,Vissers-JCP140_2014_144902}.
Structures formed from more complex patchy particles have also been  
studied~\cite{Chen-Nature469_2011_381,Chan-Science331_2011_199,Chen-Langmuir28_2012_13555,%
Romano-jpcm24_2012_064113}. 
Chen and co-workers~\cite{Chen-Nature469_2011_381}
produced tri-block patchy particles and showed the formation of a colloidal kagome lattice.
Chen's group also studied the self-assemblies formed by multiblock patchy particles~\cite{Chan-Science331_2011_199}.

For one-patch particles, 
the effects  of interaction length~\cite{Preisler-vss-jcp145_2016_064513}
and coverage of a patch area~\cite{Preisler-vmsf_solfmatter10_2014_5121,Iwashita_Scientific_reprot6_2016_27599} on 
structures formed by these particles  have already been  studied in a three-dimensional system.
In these  studies, 
the coverage of the patch area was fixed to one half of the particle's surface
to study the effect of the interaction length~\cite{Preisler-vss-jcp145_2016_064513},
and a short interaction length was set when the effect of the coverage of patch area
was studied~\cite{Preisler-vmsf_solfmatter10_2014_5121}.
If we set a longer interaction length and varied  the coverage of a patch area,
we expect various structures that have not been reported until now to be  formed even in a two-dimensional system.

In this paper, we describe how the structures formed by one-patch particles
change  with increasing the coverage of patch region when the attraction is long-ranged.
We expect that the structures formed in two--dimensional systems are simpler than those in three-dimensional systems.
Studying such structures is important because 
two-dimensional regular structures are used as   in the colloidal epitaxy method~\cite{Blaaderen}
to form regular three-dimensional structures.
\color{black}
In Ref.~\onlinecite{Iwashita_Scientific_reprot6_2016_27599},
Iwashita et al. observed the orientational order of patch area and the positional order in thin system
for short-range interaction. 
The authors showed  that  the orientational order  of patch area and the positional order  changes intricately
with increasing the thickness of the layers.
Studying these orders in two-dimensional systems is important to 
understand  how the orders in the two-dimensional system changes with increasing the the thickness of the system
and how the orders in two-dimensional system are related to those in the three-dimensional system.
\color{black}
Therefore, as a  first step, we study the two-dimensional structures formed by spherical one-patch particles.
The orientational ordering of the patch direction has already been studied 
assuming  that the hexagonal lattice is formed in two-dimensional systems~\cite{
Shin-Schweizer_softmatter10_2014_262,Iwashita-k_softmatter10_2014_7170}.
Studying the orientational order under this  assumption is probably reasonable
for short-range interactions when the pressure is high.
However, if the attractive interaction between patchy particles  is long and the pressure is low,
it is not obvious whether the hexagonal lattice is formed or not.
Thus, 
to study which structures are produced and how the  patchy particles are oriented, 
we perform isobaric-isothermal (NPT) Monte Carlo simulations using 
the  Kern--Frenkel (KF) potential~\cite{Kern-f-jcp118_9882_2003}.
In Sec.~\ref{sec:model-npt}, we introduce our model used in NPT simulations
and in Sec.~\ref{sec:results-npt}, we present the results of the NPT  simulations.
In the section, we briefly discuss each result
and provide a summary of our results in Sec.~\ref{sec:summary}.

\section{Model for Monte Carlo simulation}\label{sec:model-npt}

\begin{figure}[htp]
\centering

\includegraphics[width=10.0cm,clip]{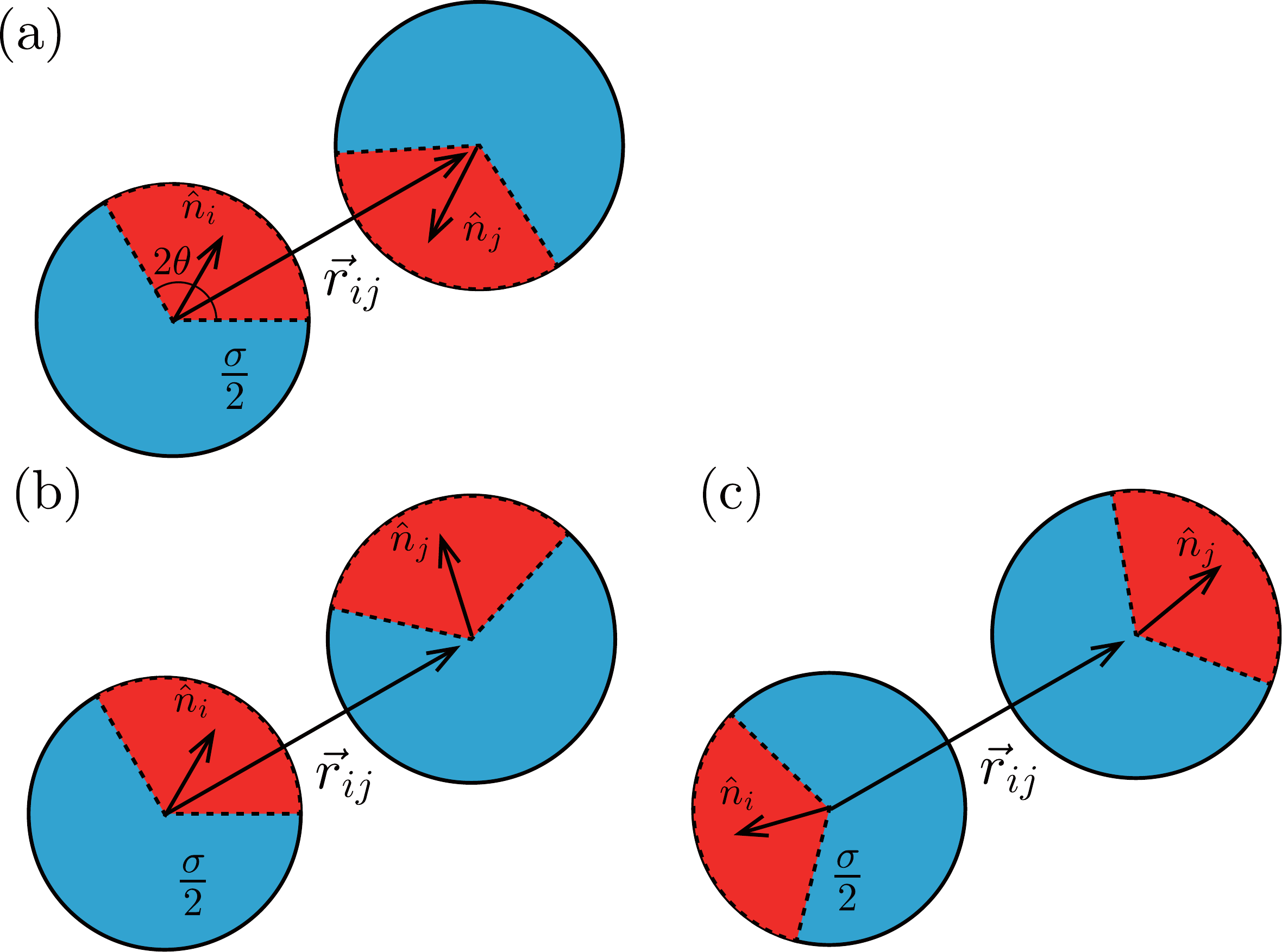} 

\caption{
(color online)
Interaction between patchy particles
for which the  diameter and attractive range  are $\sigma$ and  $\Delta /2$, respectively.
Patch-facing particles attract each other 
when their  patch areas [red(dark)  areas] satisfy the conditions,
$\hat{\bm{n}}_i \cdot \bm{r}_{ij}/|\bm{r}_{ij}| > \cos \theta$  and 
$\hat{\bm{n}}_j \cdot \bm{r}_{ji}/|\bm{r}_{ij}| > \cos \theta$,
as in (a),
whereas 
in (b) and~(c),
their interaction is simply hard-core  repulsive.
}
\label{fig:model}
\end{figure}

We performed NPT Monte Carlo simulations to study stable structures under given temperatures and 
pressures. In the simulations, we used the KF potential~\cite{Kern-f-jcp118_9882_2003}
as an attractive potential between particles.
We assume that spherical particles  have a one-patch region.
For the KF potential~\cite{Kern-f-jcp118_9882_2003},
the interaction  potential of the $i$th and $j$th particles
is expressed as  
\begin{equation}
U_\mathrm{KF} (\bm{r}_{ij},\hat{\bm{n}}_{i}, \hat{\bm{n}}_{j} ) 
= U_\mathrm{rep}(r_{ij}) 
+ U_\mathrm{att}(r_{ij}) f (\bm{r}_{ij}, \hat{\bm{n}}_{i}, \hat{\bm{n}}_{j} ),
\label{eq:KFpotential}
\end{equation}
where 
$\bm{r}_i$ denotes  the center of mass for the $i$th  particle,
$\bm{r}_{ij} = \bm{r}_j-\bm{r}_i$,
$r_{ij} =|\bm{r}_{ij}|$,
and \color{black} $\hat{\bm{n}}_i=(n_{ix}, n_{iy}, n_{iz})$
\color{black} represents the direction of the patch region
of the $i$th particle.
The first term $U_\mathrm{rep}(r_{ij})$ represents a hard-core repulsive potential, which is given by 
\begin{equation}
U_\mathrm{rep}(r_{ij})
=
 \begin{cases} 
  \infty    & (r_{ij} \le  \sigma) \\
  0 &  (\sigma  < r_{ij})
 \end{cases},
\label{eq:hard-core}
\end{equation}
where $\sigma $ is the diameter of the patchy particles.
The second term in Eq.~(\ref{eq:KFpotential}) represents the attractive part of KF potential.
$U_\mathrm{att}(r_{ij})$ is the  square-well potential given by 
\begin{equation}
U_\mathrm{att}(r_{ij})
=
 \begin{cases} 
  -\epsilon      & (\sigma < r_{ij} \le  \sigma + \Delta ) \\
  0                 &  (\sigma + \Delta  < r_{ij})
 \end{cases},
\label{eq:square-well}
\end{equation}
where $\epsilon$ is a positive parameter  representing  the well depth
and $\Delta/2 $ is the attraction range for each particle.
 $f (\bm{r}_{ij}, \hat{\bm{n}}_{i}, \hat{\bm{n}}_{j} )$ describes 
how the attraction depends on the patch directions of the $i$th and $j$th particles
and is given by 
\begin{equation}
f (\bm{r}_{ij}, \hat{\bm{n}}_{i}, \hat{\bm{n}}_{j} )
=
 \begin{cases} 
  1 & {\color{black} \text{$(\hat{\bm{n}}_{i} \cdot \bm{r}_{ij}/|\bm{r}_{ij}|  > \cos \theta $ and 
        $ \hat{\bm{n}}_{j} \cdot \bm{r}_{ji}/ |\bm{r}_{ji}| > \cos \theta $ })} \\
  0 & \text{otherwise}
 \end{cases}.
\label{eq:ftheta}
\end{equation}
where $\theta $ is related to the ratio of  the patch region  to the periphery $\chi$;
specifically, $\chi = (1-\cos \theta)/2$.
Fig.~\ref{fig:model} shows the interaction given by the KF potential.
Patch-facing particles attract each other [Fig.~\ref{fig:model}(a)], 
otherwise the interaction between particles is repulsive [Figs.~\ref{fig:model}(b) and (c)].

To study how $\chi$ affects  two-dimensional structures formed by patch particles,
we perform NPT Monte Carlo simulations.
In the simulations, we set   $\Delta $  to $\sigma/2$ and the number of particles $N$ to $256$.
We consider a square system for which the size is $L \times L$.
Initially, we place $N$ patchy particles in the system at random.
We move the particles for a long time neglecting the second term of  $U_\mathrm{KF}$ to remove the effect of 
the initial configuration.
Then,
we take the attraction term into  account and 
perform translational trials, rotational trials, and trials in which the system size is changed.
We tune up  the absolute values of translation, rotation, and change in system size 
to maintain their  acceptance ratios above $0.3$.
For simplicity, 
instead of checking the Gibbs free energy to assess whether the system has reached 
an equilibrium state,
we monitored the system size and its internal energy.
When they seem to be saturated, we consider that the system reaches to an equilibrium state.

\section{Results of Monte Carlo simulation}\label{sec:results-npt}

We performed Monte Carlo simulations with some values of $\theta$
to study how pressure affects the two-dimensional structures for each $\theta$
and to identify differences from structures formed by  a short-range attraction.
\color{black}
First, we show typical structures for some values of $\theta$.
Then, we show the phase diagram
for some $\epsilon$. Finally, we consider how the structures change with the form of attraction.
\color{black}

\subsection{Structures for $\theta =15^\circ$}

\begin{figure}[htp]
\centering

\includegraphics[width=7.0cm,clip]{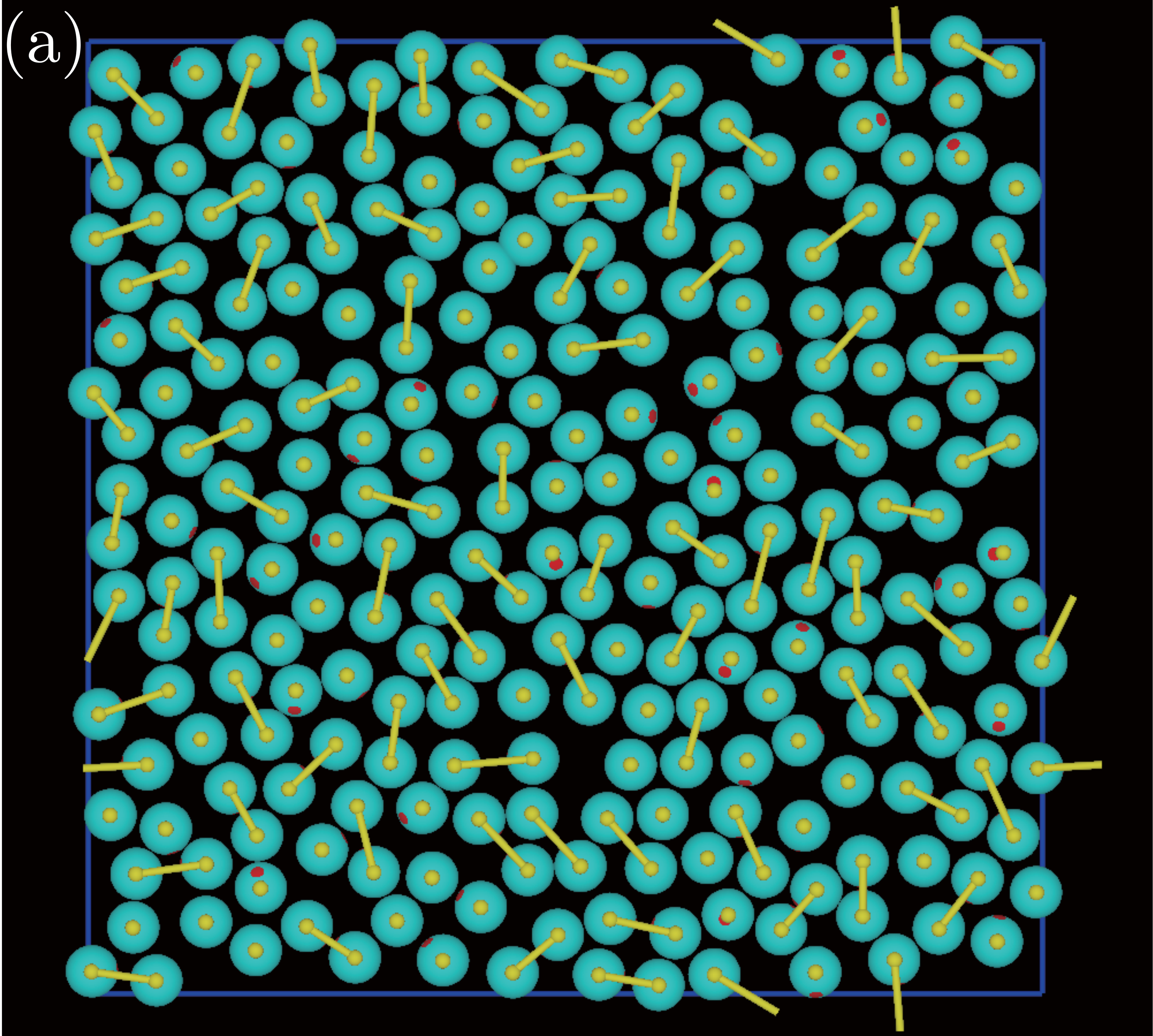} 
\includegraphics[width=7.0cm,clip]{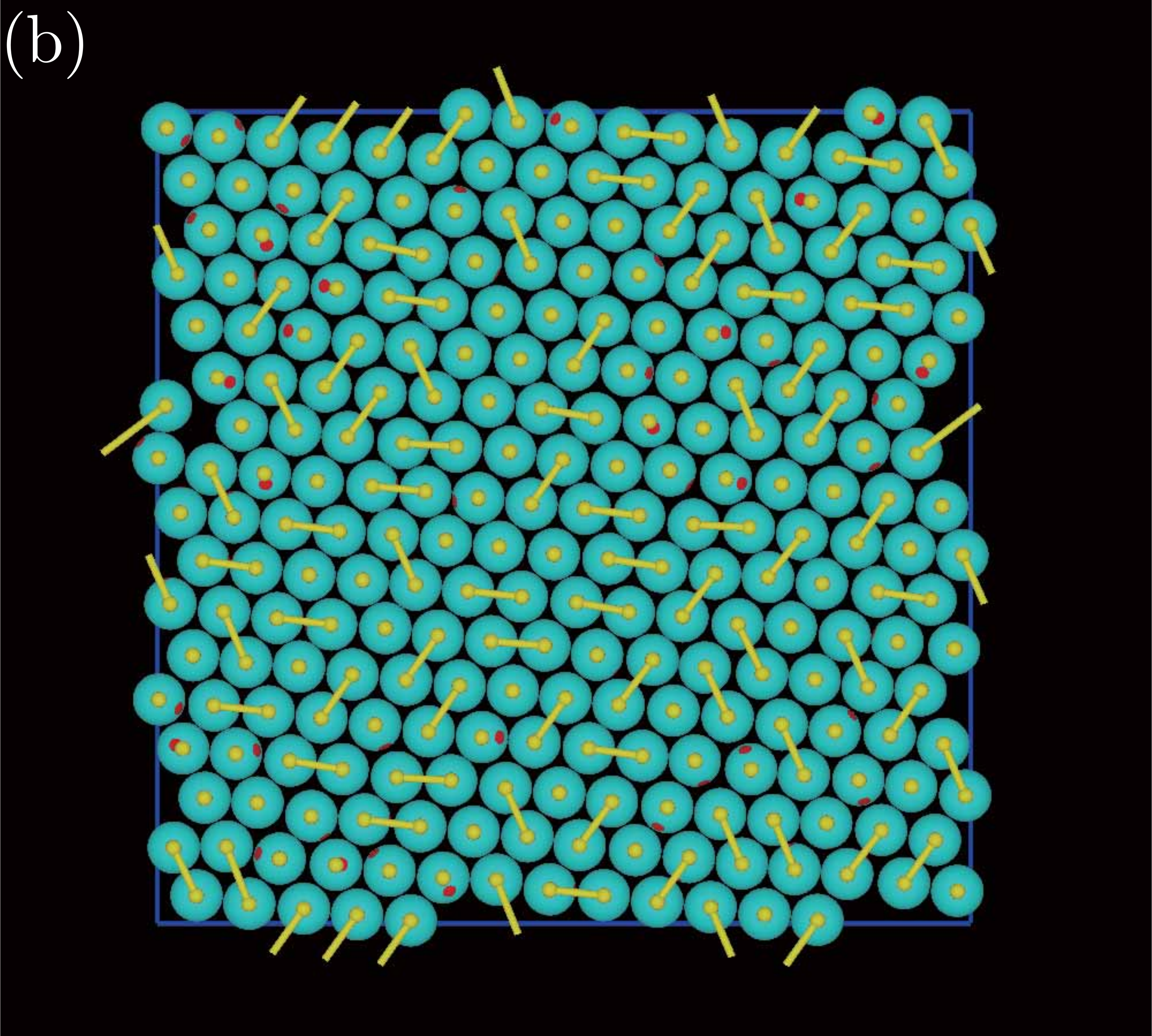}

\caption{
(color online)
Snapshots of two-dimensional structures  with $\theta =15^\circ$ at  (a) $P\sigma^3/k_\mathrm{B}T=5.0$
and (b) $P\sigma^3/k_\mathrm{B}T=40$.
The red(dark) regions represent the patch areas of attractive interaction
between particles under the KF potential;
yellow(light) lines mark the connections between particles.
}
\label{fig:theta=15}
\end{figure}
Figure~\ref{fig:theta=15} presents snapshots of the structure with $\theta=15^\circ$.
The scaled pressures $P\sigma^3/k_\mathrm{B}T$ are $5$ 
for  Fig.~\ref{fig:theta=15}(a) and $40$ for  Fig.~\ref{fig:theta=15}(b).
In these figures, areas generating an attraction between patchy particles
are marked in red;
We put small yellow spheres at the centers of particles
and draw yellow lines between attracting particles.
Dimers of patchy particles are formed in the case of low pressure [Fig.~\ref{fig:theta=15}(a)].
When the pressure is higher [Fig.~\ref{fig:theta=15}(b)], the dimers are arranged to form a hexagonal lattice, 
in which the directions  of dimers seem to be at random.

\begin{figure}[htp]
\centering

\includegraphics[width=10.0cm,clip]{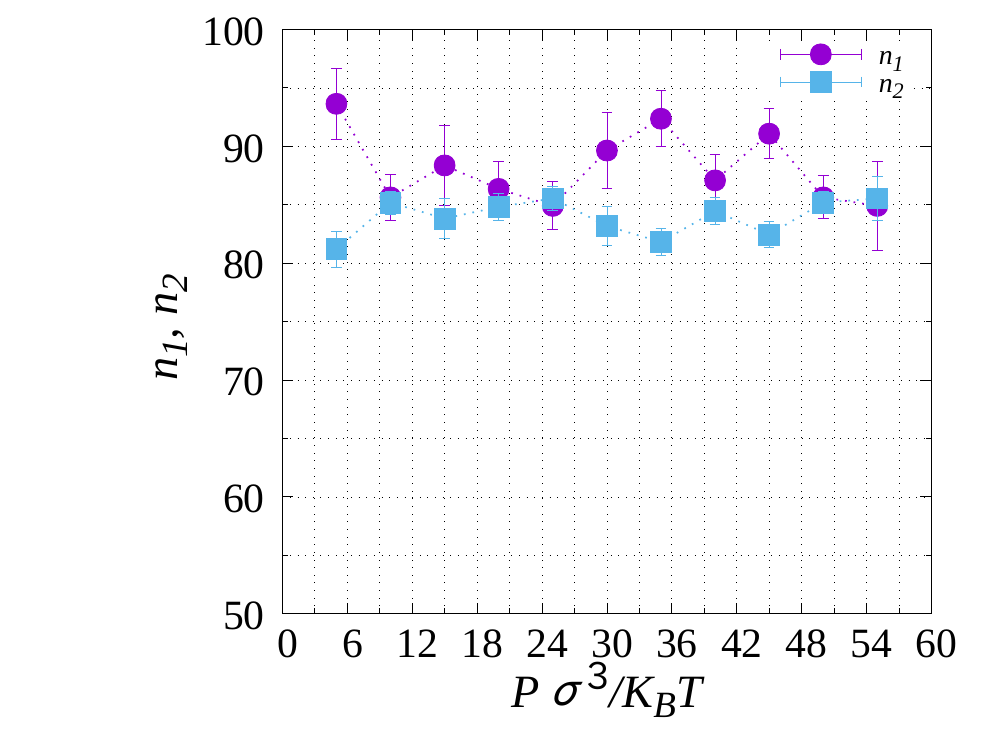} 

\caption{
(color online)
Dependence of $n_1$ and $n_2$ on pressure for  $\theta = 15^\circ$,
where $n_1$ and $n_2$ represent the numbers of monomers  and dimers, respectively.
In the later stage of the simulations, 
data are collected and averaged over 10 points every $4 \times 10^5$ MC steps.
}
\label{fig:theta=15a}
\end{figure}
In Fig.~\ref{fig:theta=15},
clusters larger than dimers are not observed,
probably because $\chi$ is too small to make large clusters.
Figure~\ref{fig:theta=15a} shows how the number of monomers $n_1$ and that of dimers  $n_2$ depend on pressure.
Because it is reasonable  to believe that the system at two different time step in a run are independent
if observed  at sufficiently  long Monte-Carlo step intervals,
we averaged the data over 10 points over an interval of $4 \times 10^5$ Monte Carlo steps. 
Taking into account that the number of particles is  not so large in our simulations,
fluctuations in $n_1$ and $n_2$ are inevitable. 
These numbers appear independent of pressure and roughly the same.

\subsection{Structures for $\theta =30^\circ$}

\begin{figure}[htp]
\centering

\includegraphics[width=7.0cm,clip]{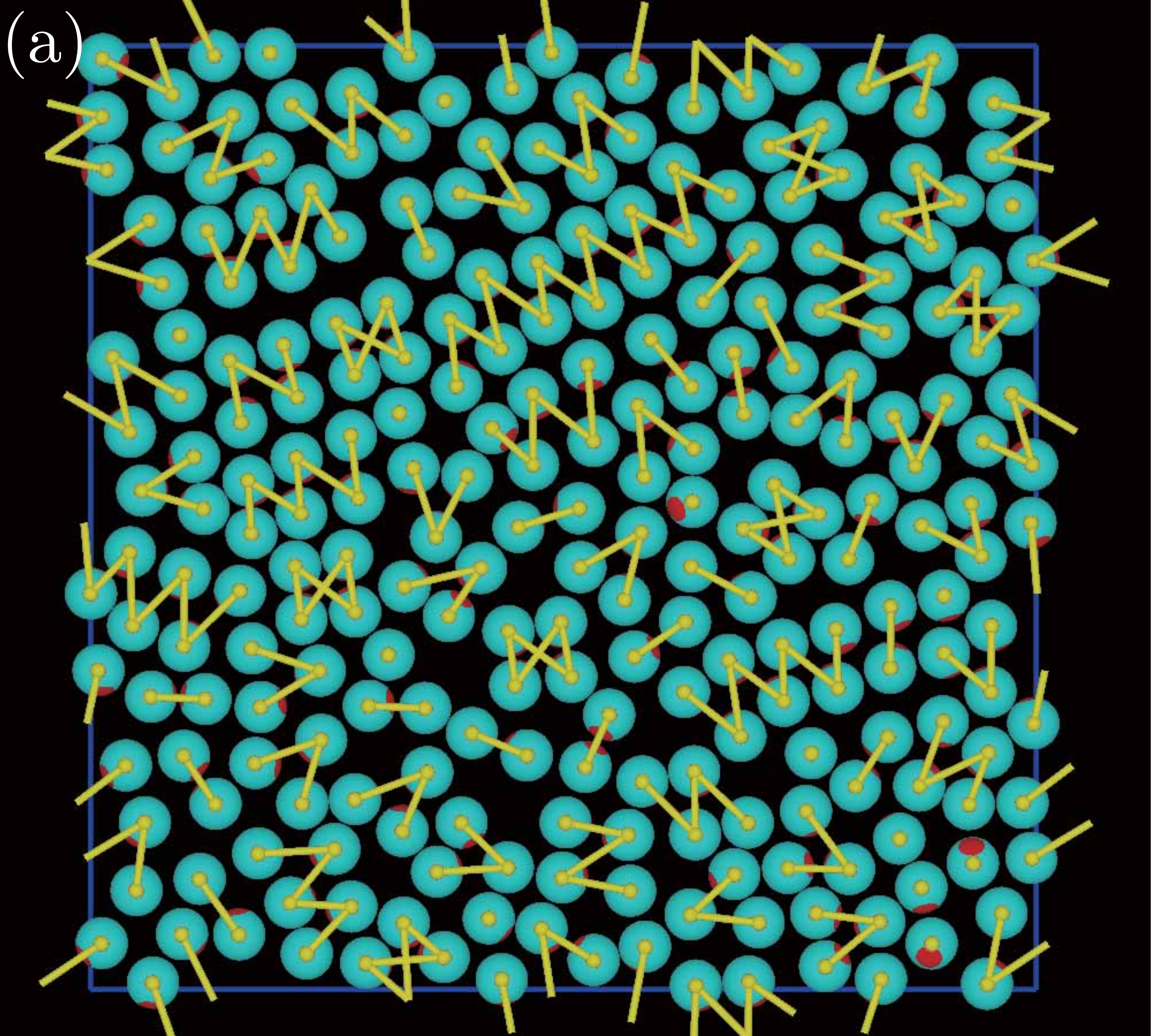} 
\includegraphics[width=7.0cm,clip]{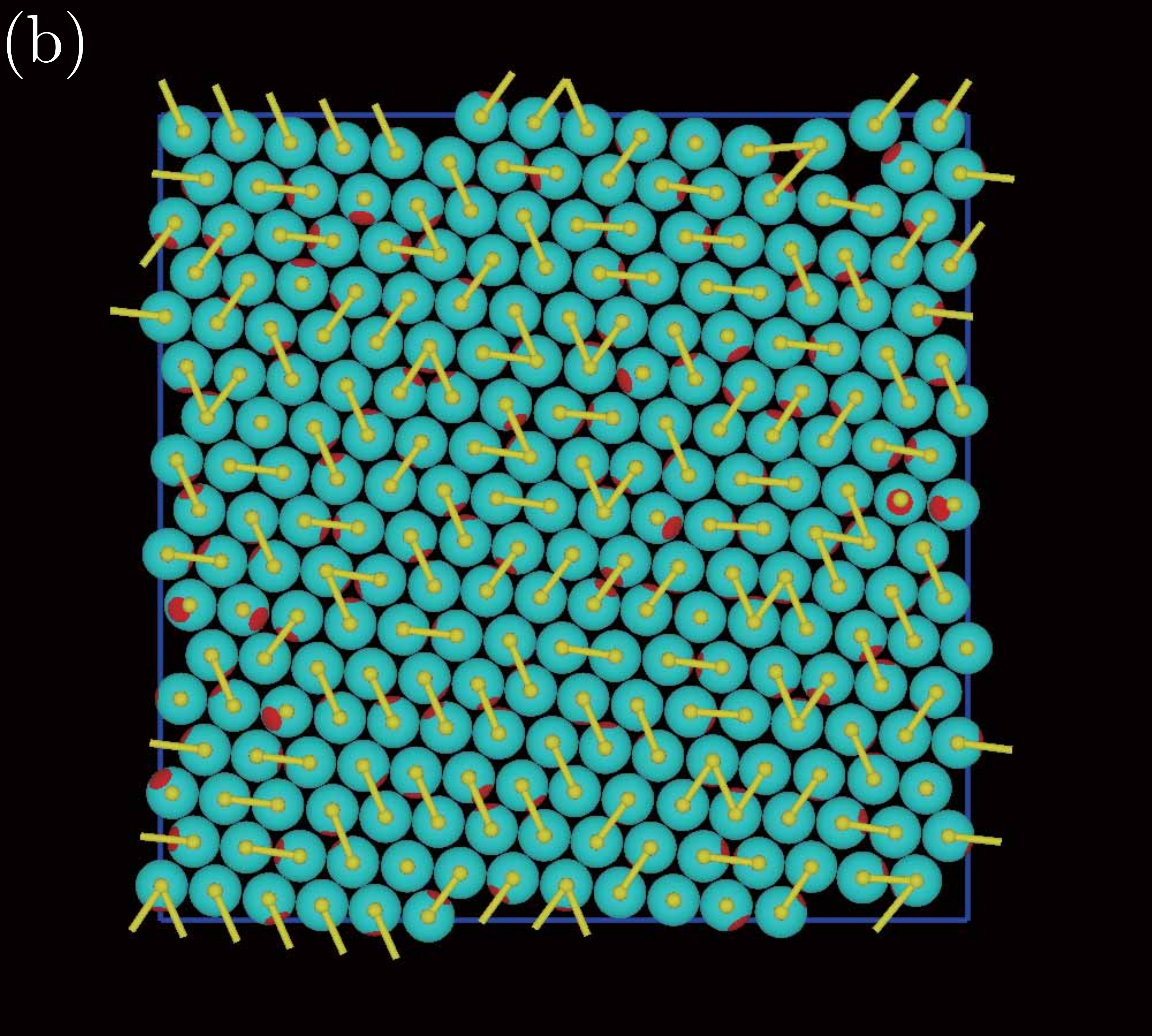}

\caption{
(color online)
Snapshots of two-dimensional structures  with $\theta =30^\circ$ at  (a) $P\sigma^3/k_\mathrm{B}T=5.0$
and (b) $P\sigma^3/k_\mathrm{B}T=55$.
The significance of the red(dark) regions and 
yellow(light) lines is the same as given by in Fig.~\ref{fig:theta=15}.
}
\label{fig:theta=30}
\end{figure}
Figure~\ref{fig:theta=30} shows snapshots for $\theta=30^\circ$.
The lattice structure formed under this high pressure is the hexagonal lattice consisting of dimers,
which is the same as  that formed for $\theta=15^\circ$ [Fig.~\ref{fig:theta=15}(b)].
The clusters organized in low pressure  [Fig.~\ref{fig:theta=30}(a)]  are different from the dimers shown in Fig.~\ref{fig:theta=15}(a);
zigzag chains of patchy particles have formed under loose attractions as well as  compact square tetramers.
When $\theta=30^\circ$ 
and 
the interaction length is short enough that the attraction acts between contacting particles,  
the conditions \color{black}  $\hat{\bm{n}}_{i} \cdot \bm{r}_{ij}/|\bm{r}_{ij}| > \cos \theta $
and  $ \hat{\bm{n}}_{j} \cdot \bm{r}_{ji}/|\bm{r}_{ij}| > \cos \theta $ \color{black}  may not be satisfied in  these structures. 
However, 
because the attraction range is sufficiently long that distant particles attract each other, 
the angle conditions are satisfied and both zigzag chains and compact square tetramers are formed.
In particular,
patchy particles in the diagonal positions in the compact square tetramers can attract each other
because the attraction length is set to be longer than $\sqrt{2}\sigma/2$.
Note that in this instance particles in  the compact square tetramers do not attract
both neighbors at the same time
that angle conditions  is only satisfied for one or other of the neighboring  particles.

\begin{figure}[htp]
\centering

\includegraphics[width=10.0cm,clip]{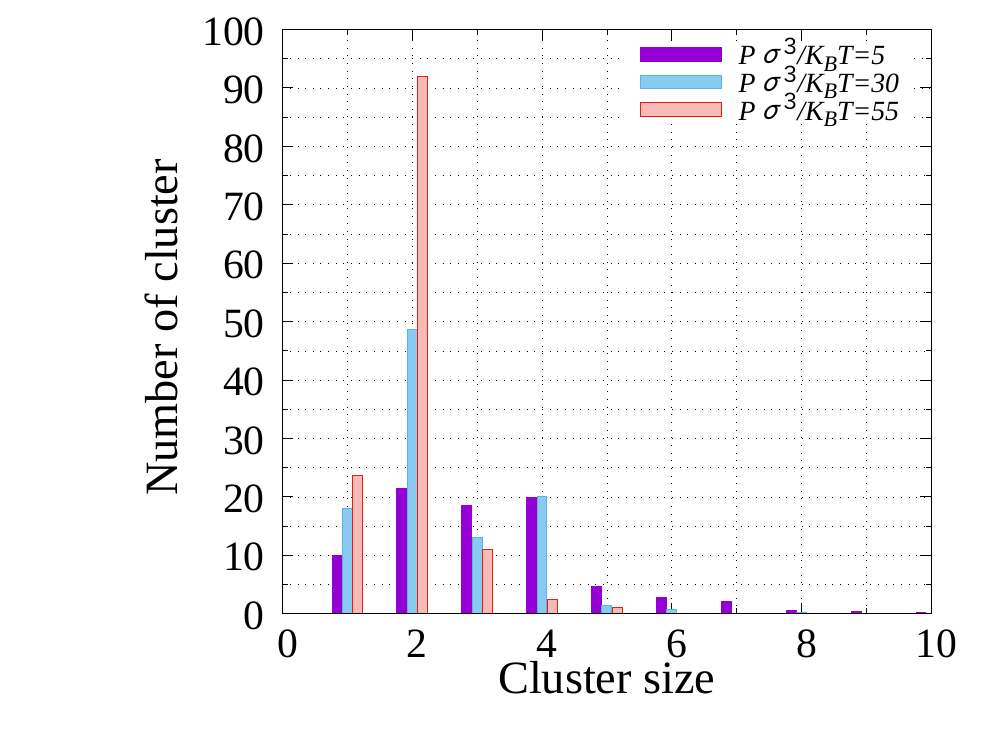} 

\caption{
(color online)
Distributions of cluster size at $\theta=30^\circ$
when $P\sigma^3/k_\mathrm{B}T=5, 30$, and $55$.
The data are averaged over 10 points every $4 \times 10^5$ MC steps in a late stage.
A few numbers of clusters  for which the size  is larger than 10 are also formed, but the numbers are negligibly small.
}
\label{fig:theta=30a}
\end{figure}
Figure~\ref{fig:theta=30a} presents
the  distributions of  cluster size  at $\theta=30^\circ$
for $P\sigma^3/k_\mathrm{B}T=5, 30$, and $55$.
A few  clusters consisting of more than 10 particles are formed in our simulations.
However,  the numbers in those clusters  are negligibly small,
and hence we only show the data for cluster sizes smaller than 10.
Hereafter, we express the number of clusters having $i$ particles as  $n_i$. 
When $P\sigma^3/k_\mathrm{B}T=5$, the distribution of the cluster size is broad and large clusters  are formed.
These clusters are mainly zigzag chain-like clusters.
$n_4$ is as large as $n_3$ because the number of tetramers is included in $n_4$.
When $P\sigma^3/k_\mathrm{B}T=30$,  
$n_1$ and $n_2$ increase, but with the expectation of $n_4$, $n_i$ with $i  \ge 3$ decrease.
This suggests 
that loose chain-like structures form with difficulty 
because,
 with decreasing the distance between particles under high pressure,
the angle conditions
\color{black}  $\hat{\bm{n}}_{i} \cdot \bm{r}_{ij}/|\bm{r}_{ij}| > \cos \theta $
and  $ \hat{\bm{n}}_{j} \cdot \bm{r}_{ji}/|\bm{r}_{ij}| > \cos \theta $ \color{black} are not  satisfied.
As square tetramers are compact, the effect of increasing pressure on the square tetramers  is weak 
at this  pressure and $n_4$ barely changes.
The pressure is so high that the square tetramers are broken when $P\sigma^3/k_\mathrm{B}T=55$.
Comparing with the case with $\theta=15^\circ$,
the number of dimers is much larger than that of monomers because the patch area is large enough 
for dimers to form  easily.

\subsection{Structures for $\theta =40^\circ$}

\begin{figure}[htp]
\centering
\includegraphics[width=7.0cm,clip]{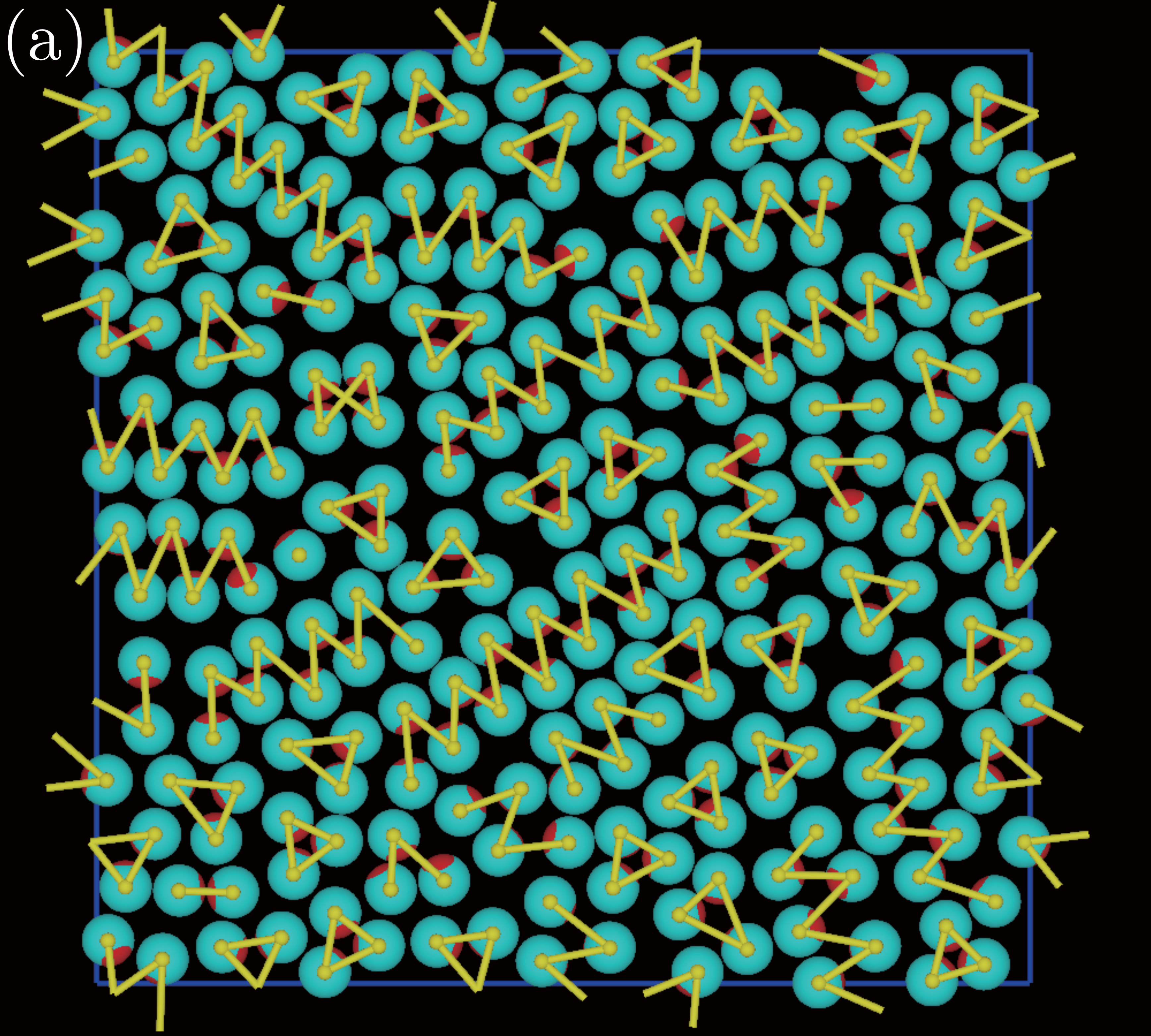} 
\includegraphics[width=7.0cm,clip]{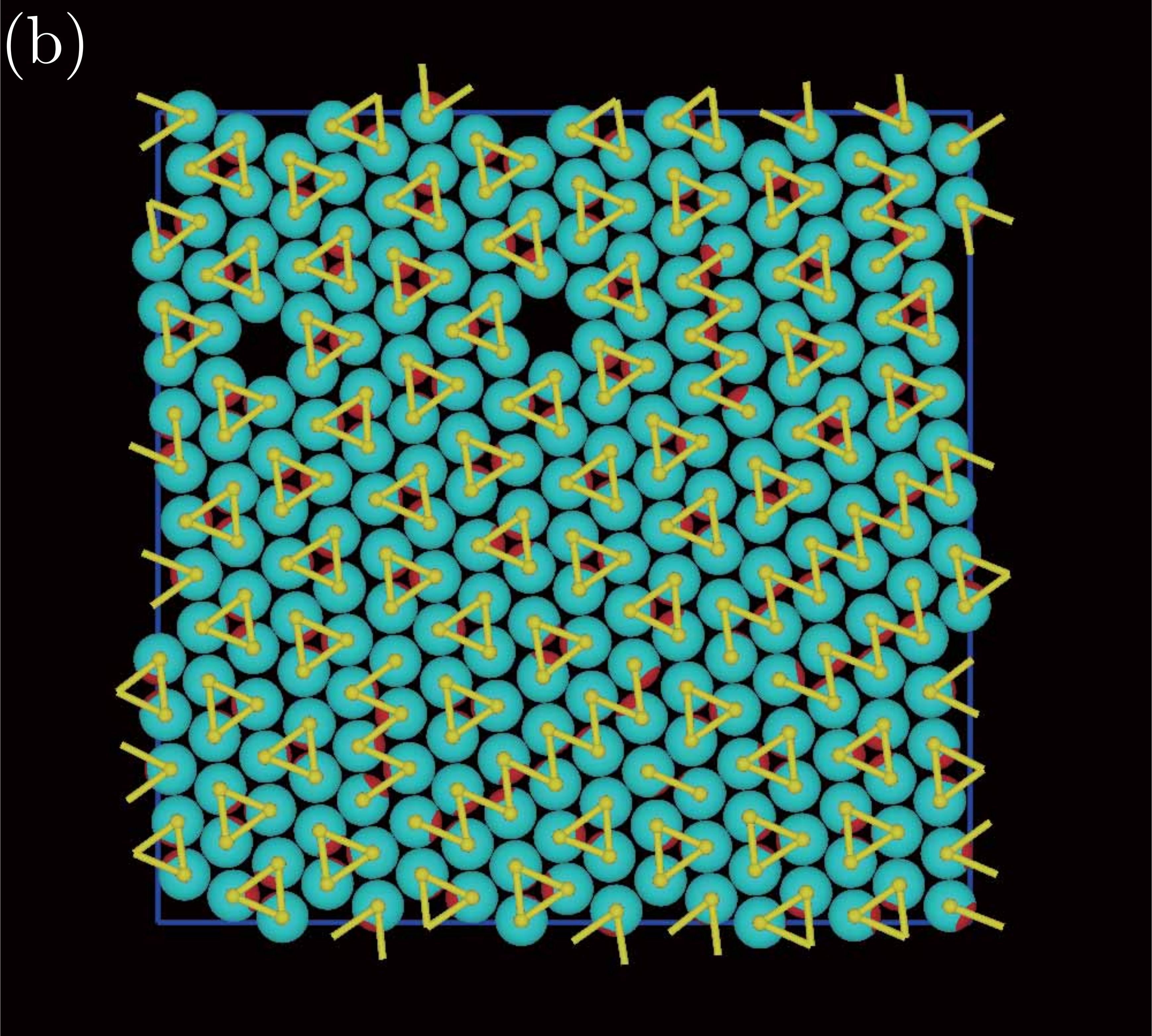}

\caption{
(color online)
Snapshots of two-dimensional structures  with $\theta =40^\circ$
with  (a)  $P\sigma^3/k_\mathrm{B}T=5$ 
and (b)   $P\sigma^3/k_\mathrm{B}T=45$. 
The significance of the red(dark) regions and 
yellow(light) lines is the same as given by in Fig.~\ref{fig:theta=15}.
}
\label{fig:theta=40}
\end{figure}
\begin{figure}[htp]
\centering

\includegraphics[width=10.0cm,clip]{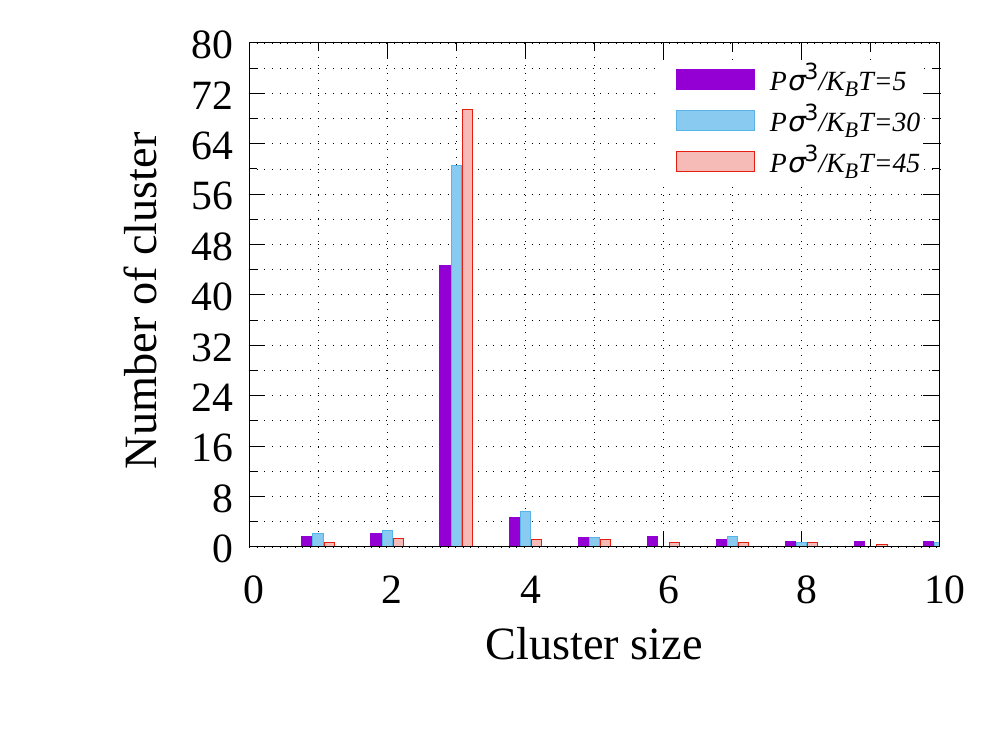} 

\caption{
(color online)
Distributions of cluster size at $\theta=40^\circ$
for $P\sigma^3/k_\mathrm{B}T=5, 30$, and $55$.
The data are collected in the late stages and averaged over 10 points every $4 \times 10^5$ MC steps.
A few numbers of clusters  for which the size  is larger than 10 are also formed but the numbers are negligibly small.
}
\label{fig:theta=40a}
\end{figure}
When $\theta=40^\circ$,  
and triangular trimers are organized under a low pressure in addition to zigzag chain-like clusters [Fig.~\ref{fig:theta=40}(a)]
because $\chi$ is large enough for particles  in the trimers to attract the other two particles. 
We observed the zigzag chain-like clusters when $\theta =30^\circ$ as well,
but triangular trimers are formed when $\theta=40^\circ$.
The number of bonds per particle in the timers is  two, which is the same as 
the bond number per a particle  in the tetramers observed in Fig.~\ref{fig:theta=30}(a).
However,  since the particle density can be higher when the triangular clusters are formed,
the trimers are preferred to tetramers to decrease the system volume.
When the pressure is high  [Fig.~\ref{fig:theta=40}(b)],
the hexagonal lattice is formed as for $\theta= 15^\circ$ and $30^\circ$.
However,  the attraction of particles is different from these two cases:
the hexagonal lattice consists of dimers for  $\theta= 15^\circ$ and $30^\circ$,
but the lattice is formed by triangular trimers for $\theta=40^\circ$.
Figure~\ref{fig:theta=40a} shows 
the  distributions of cluster size  at $\theta=40^\circ$
for $P\sigma^3/k_\mathrm{B}T=5, 30$, and $45$.
We can confirm that $n_3$ is larger than that in the cases of $\theta= 15^\circ$ and $30^\circ$,
probably as a consequence of the increase in the  number of trimers.

\subsection{Structures for $\theta =50^\circ$}

\begin{figure}[htp]
\centering

\includegraphics[width=7.0cm,clip]{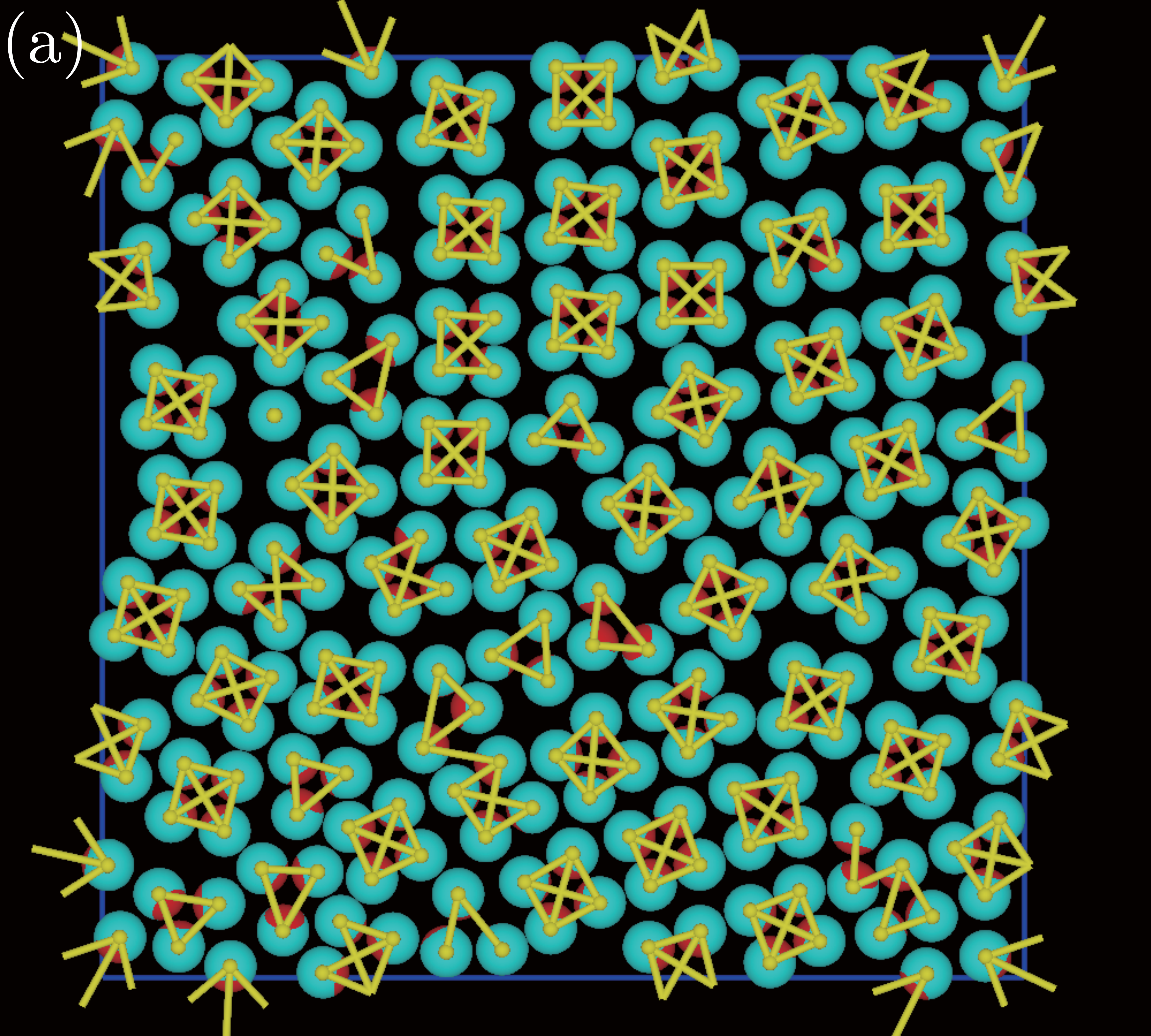} 
\includegraphics[width=7.0cm,clip]{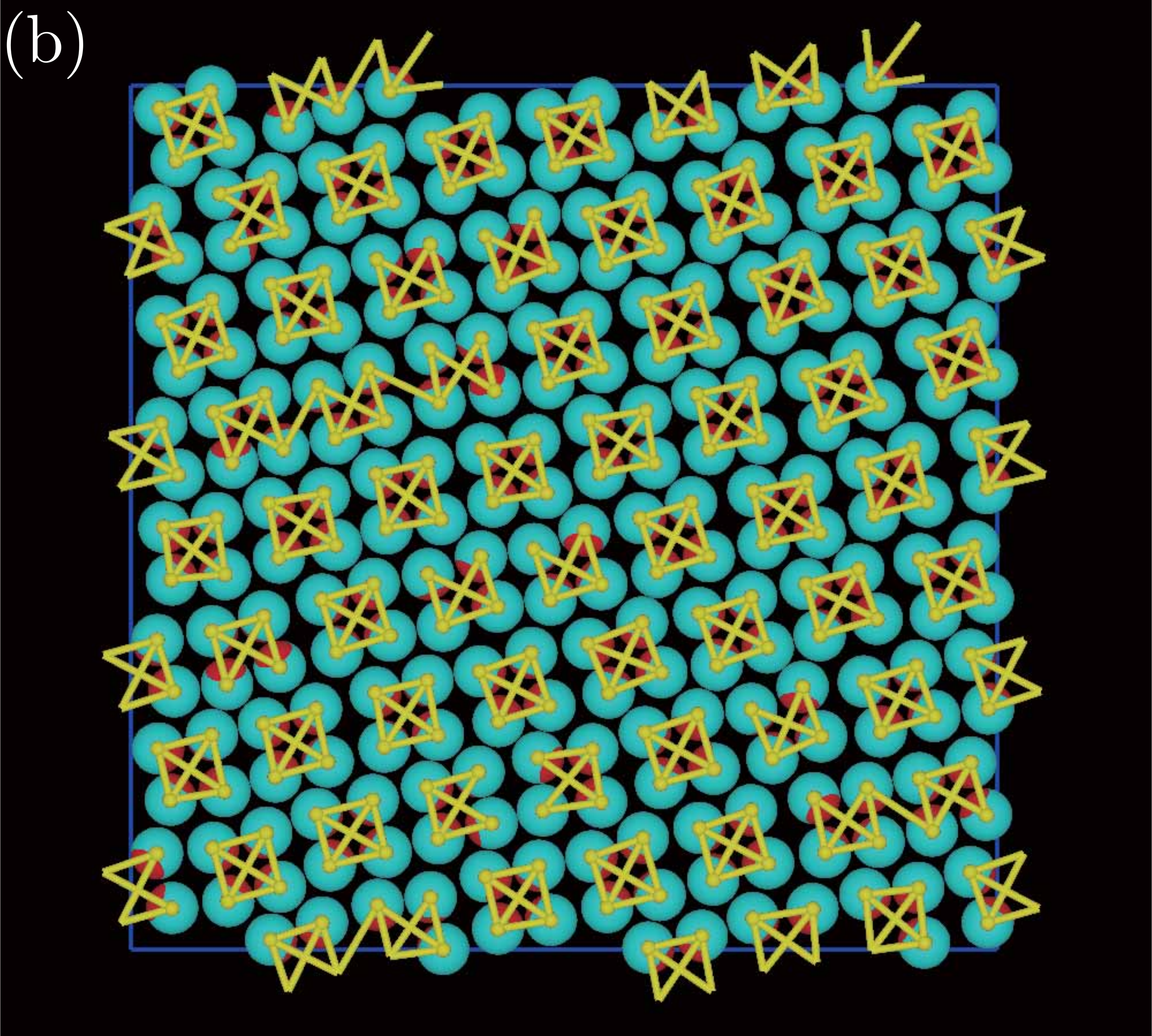}

\vspace*{0.1cm}

\includegraphics[width=7.0cm,clip]{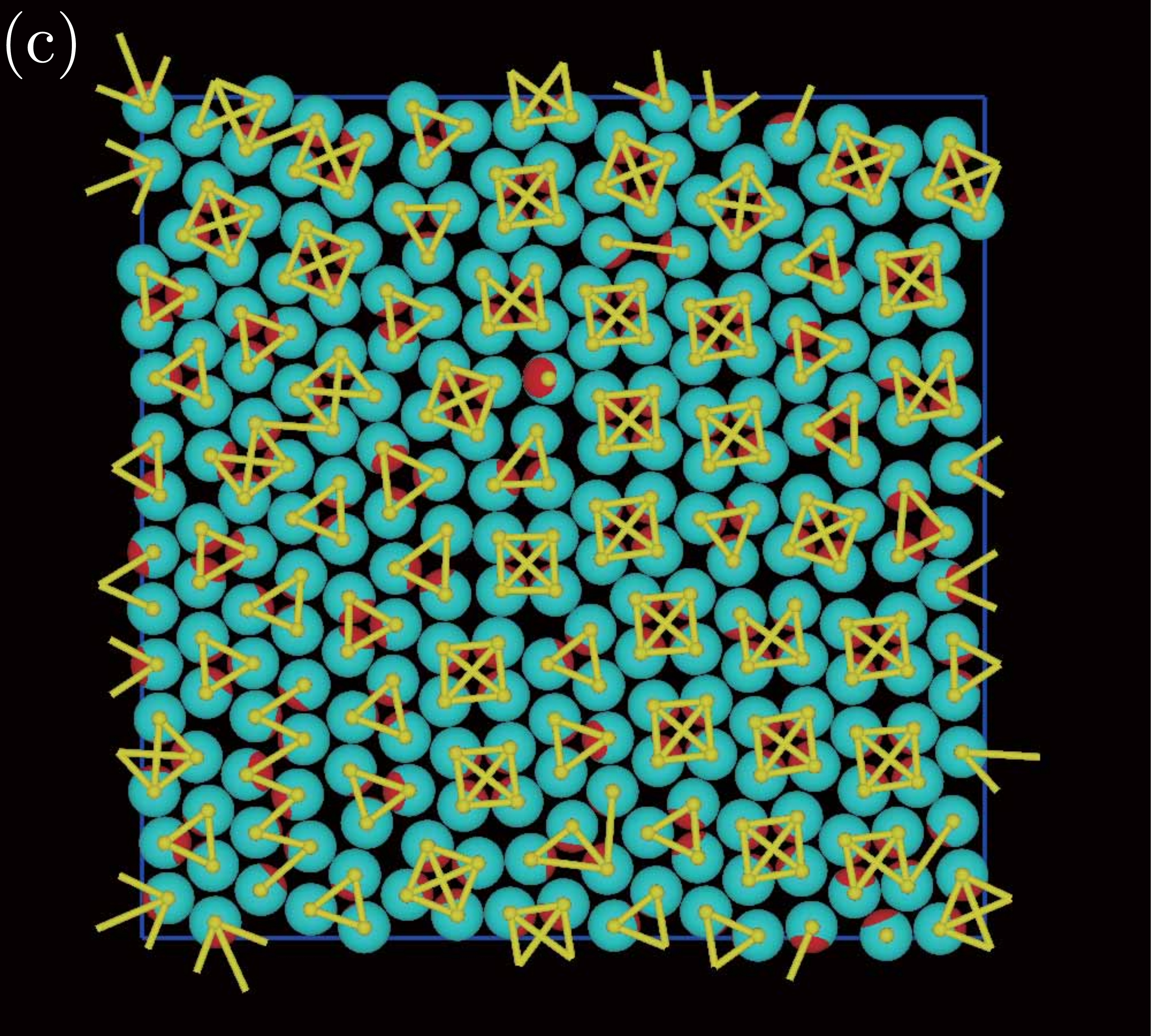} 
\includegraphics[width=7.0cm,clip]{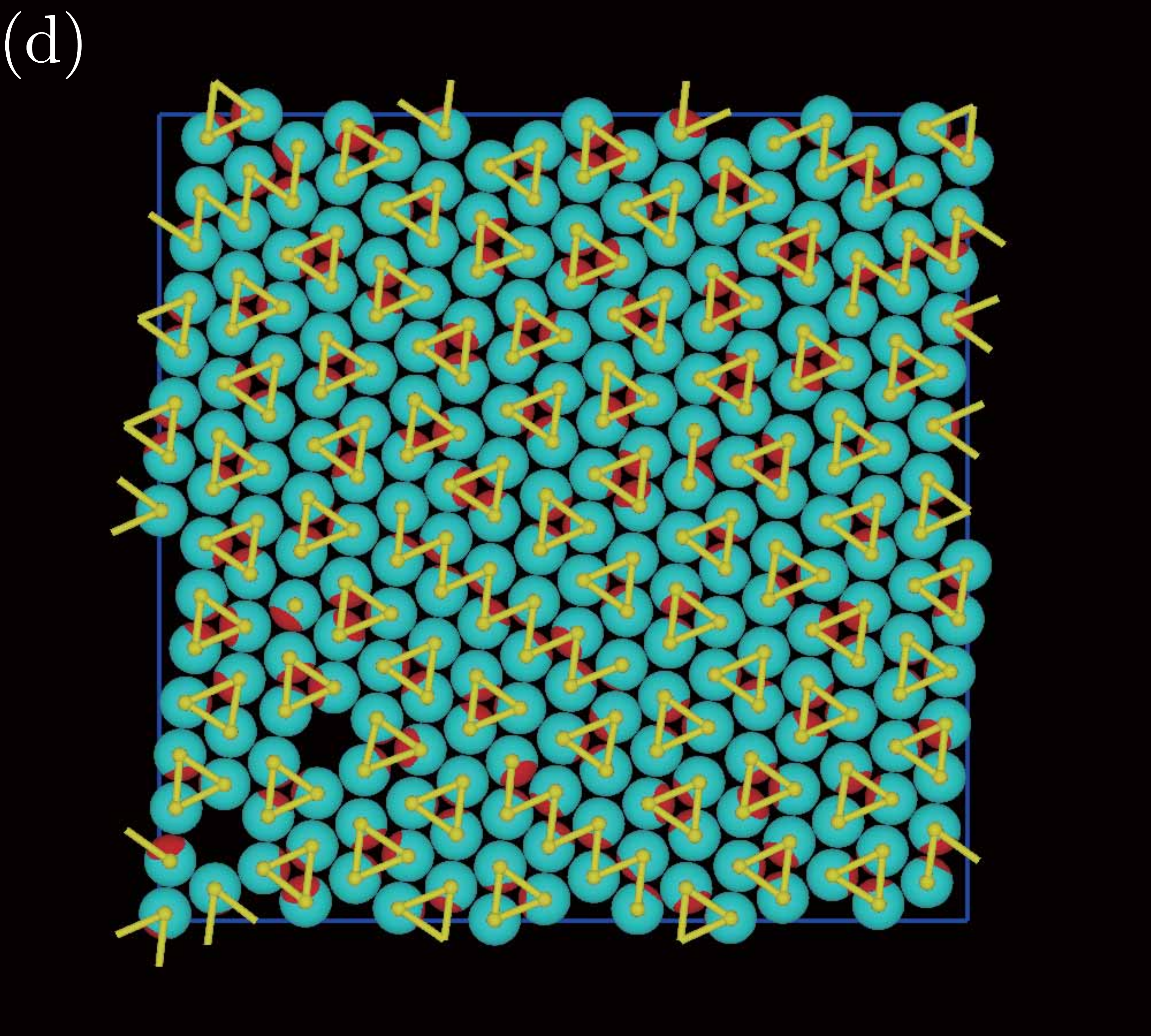}
\caption{
(color online)
Snapshots of two-dimensional structures  with $\theta =50^\circ$
with  $P\sigma^3/k_\mathrm{B}T$ equal to (a) $5$,
(b)  $10$,
(c)  $25$,
and (d)  $50$.
The significance of the red(dark) regions and 
yellow(light) lines is the same as given by in Fig.~\ref{fig:theta=15}.
}
\label{fig:theta=50}
\end{figure}
Apart from  the square tetramers and the loose zigzag chain-like clusters,
the other self-assemblies we have showed up to this point were also observed
in previous studies~\cite{Shin-Schweizer_softmatter10_2014_262,Iwashita-k_softmatter10_2014_7170}.
The effect of the long-range  attraction on the self-assemblies formed by patchy particles 
is more obvious when $\theta $ is larger than $50^\circ$.
Figure~\ref{fig:theta=50} shows snapshots for $\theta=50^\circ$.
Square tetramers are formed under low pressure [Fig.~\ref{fig:theta=50}(b)].	
Although square tetramers are also shown in Fig.~\ref{fig:theta=30}(a),
the bonding between particles in the clusters  is different;
the particles in square tetramers do not interact with one of the neighbors
when $\theta=30^\circ$,
but the particles interact  with all others when $\theta=50^\circ$ 
because $\chi$ is large [Fig.~\ref{fig:theta=50}(a)].
When $P\sigma^3/k_\mathrm{B}T=10$ [Fig.~\ref{fig:theta=50}(b)],
a regular array of the  square tetramers is formed,
which is similar to the formation of a regular array of 
supraparticles~\cite{Xia_Nat.Nanotechnol6_2011_580,Ngyyen_Proc.Natl.Acad.Sci112_2015_E3161}.

The structure is close packed with the square as basic unit.
When the pressure increases, the square tetramers fragment  to increase the particle density,
with the square tetramers and triangular trimers coexisting  [Fig.~\ref{fig:theta=50}(c)].
When  we carry out simulations with sufficiently high pressure,
a hexagonal lattice with triangular trimers forms [Fig.~\ref{fig:theta=50}(d)]. 

\begin{figure}[htp]
\centering

\includegraphics[width=10.0cm,clip]{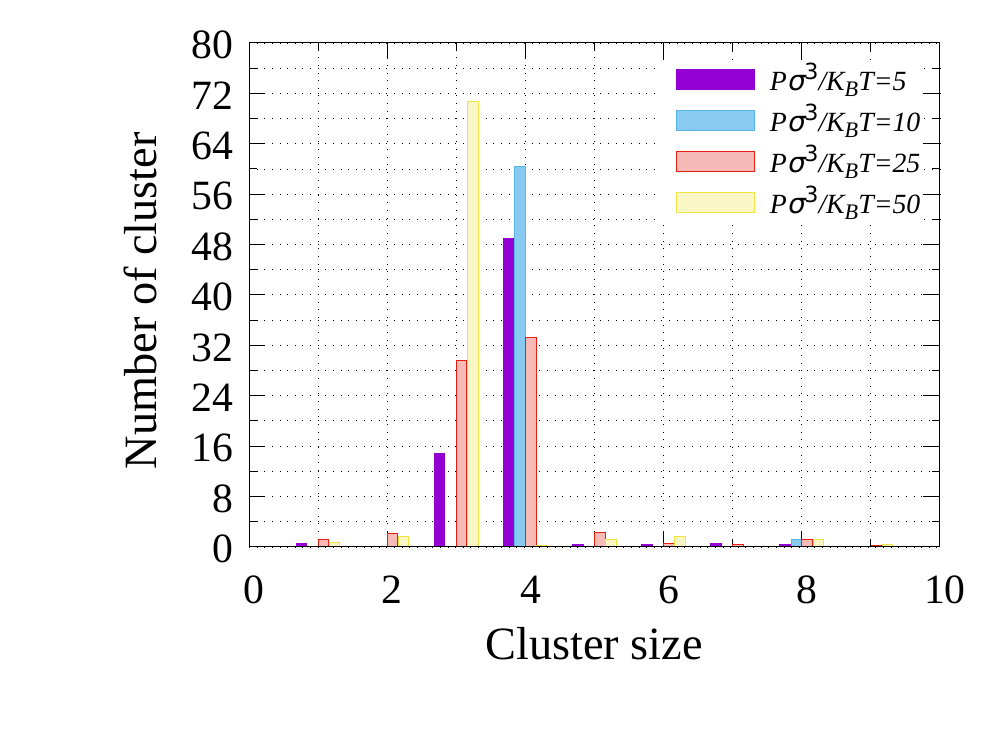} 

\caption{
(color online)
Distributions of cluster size at $\theta=50^\circ$
for $P\sigma^3/k_\mathrm{B}T=5, 10, 25$ and $55$.
Data are collected in a late stage and averaged over 10 points every  $4 \times 10^5$ MC steps.
Small numbers of clusters for which the size  is larger than 10 are also formed, but the numbers are negligibly small.
}
\label{fig:theta=50a}
\end{figure}
We show  the  distributions of cluster size  at $\theta=50^\circ$ 
for  $P\sigma^3/k_\mathrm{B}T=5, 10, 25$, and $50$ in Fig.~\ref{fig:theta=50a}.
When $P\sigma^3/k_\mathrm{B}T=5$,  
$n_4$ is larger than $n_3$, which means that a small number of triangular tetramers 
coexist with a large number of square tetramers.
When $P\sigma^3/k_\mathrm{B}T=10$, $n_3=0$ and $n_4$ increases,
indicating that triangular tetramers are eliminated from the system and almost all clusters are square tetramers. 
When $P\sigma^3/k_\mathrm{B}T=25$, $n_3$ increases again,
because the triangular trimers form  again to increase the density. 
With the formation of a large peak at $n_3$,
we believe that most of the clusters become triangular trimers when  $P\sigma^3/k_\mathrm{B}T=55$.

\begin{figure}[htp]
\centering

\includegraphics[width=10.0cm,clip]{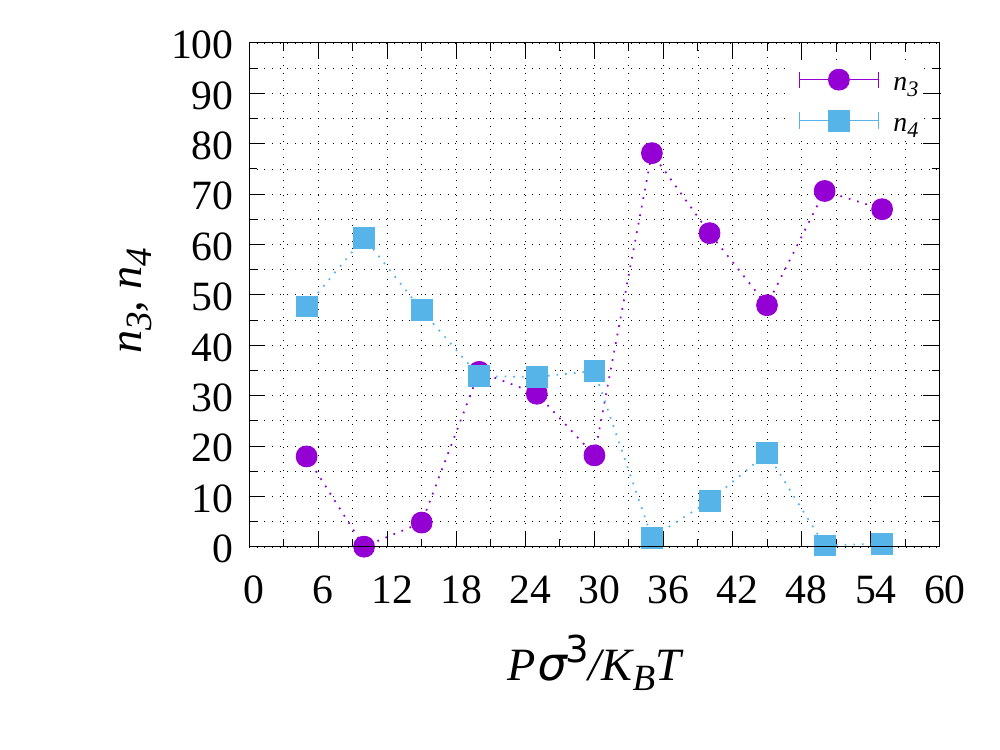} 

\caption{
(color online)
Dependence of $n_3$ and $n_3$ on pressure at $\theta = 50^\circ$,
where $n_3$ and $n_4$ represent the numbers of trimers and tetramers, respectively.
The data are collected in a late stage and averaged over 10 points every $4 \times 10^5$ MC steps.
}
\label{fig:theta=50n}
\end{figure}
When $\theta=50^\circ$, 
the structural  change induced by increasing the pressure mainly
occurs of the transition between triangular trimers and square tetramers.
Figure~\ref{fig:theta=50n} shows the dependence of $n_3$ and $n_4$  on pressure.
Unfortunately, the detail dependence is not obvious because of large fluctuations in the data
caused by the small number of clusters.
Nevertheless,
we can find $n_3$ increases and $n_4$ decreases with increasing pressure.

\subsection{Structures for $\theta =80^\circ$}

\begin{figure}[htp]
\centering
\includegraphics[width=7.0cm,clip]{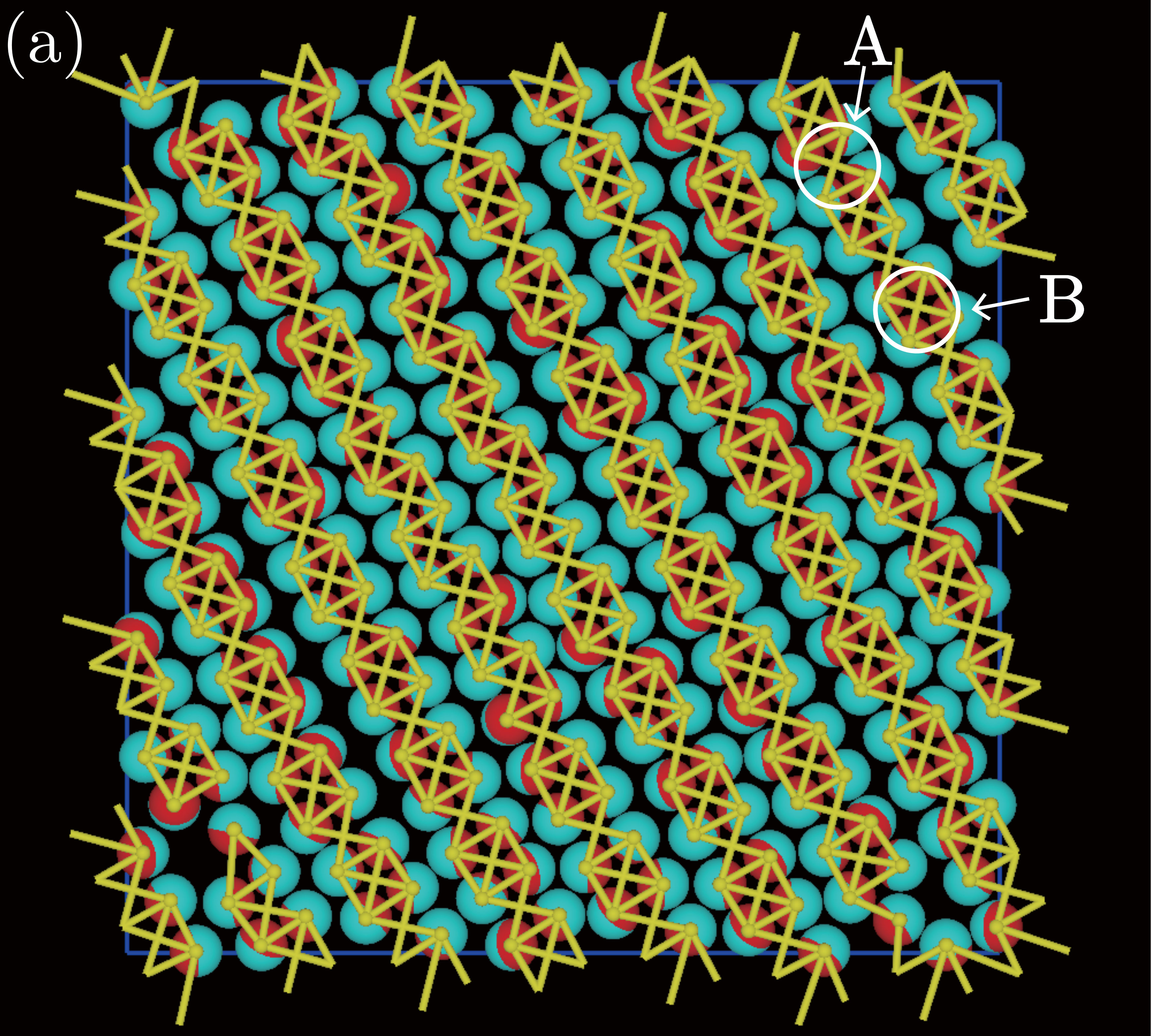} 
\includegraphics[width=7.0cm,clip]{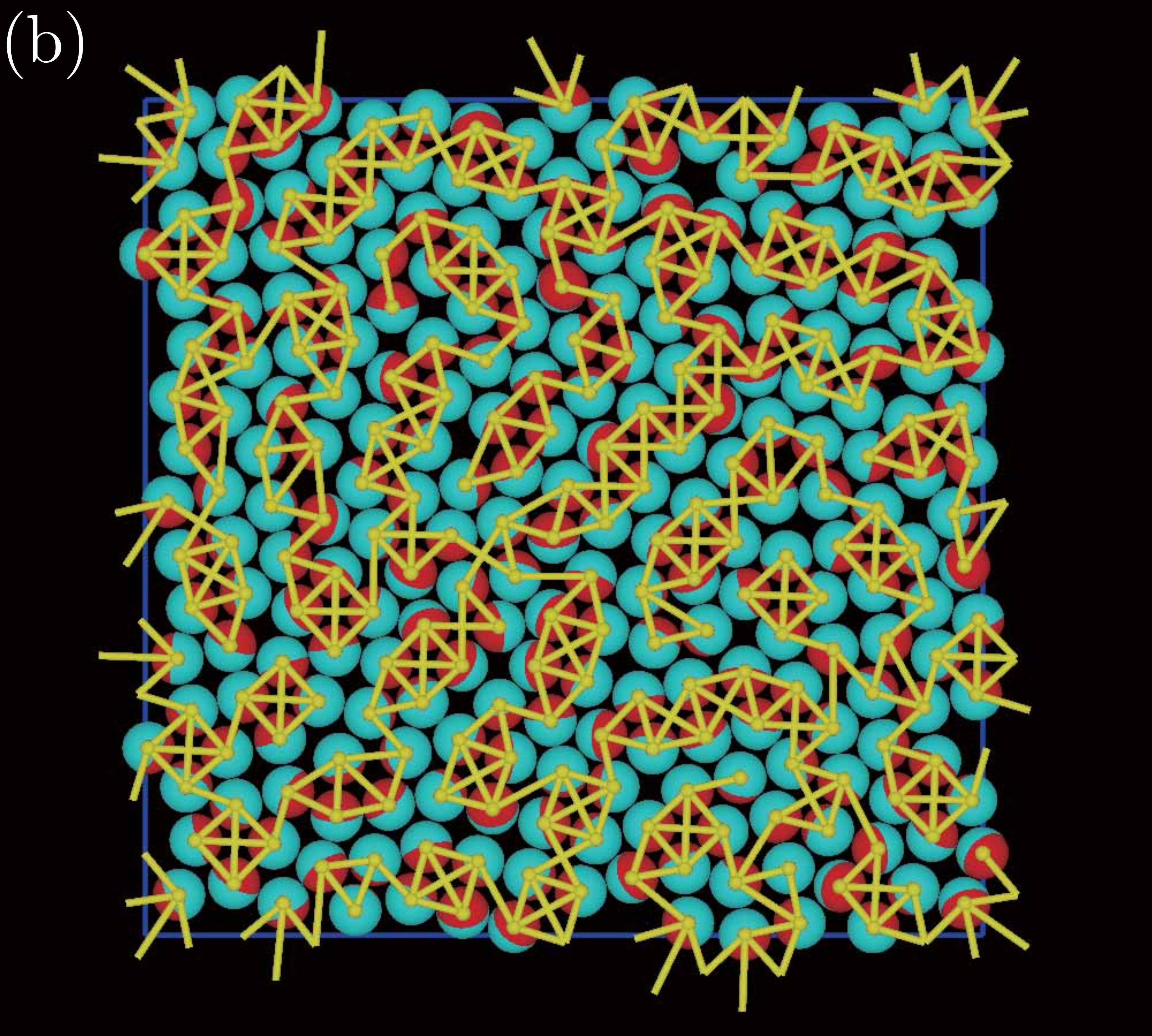}
\caption{
(color online)
Snapshots of two-dimensional structures  with $\theta =80^\circ$
with $P\sigma^3/k_\mathrm{B}T$ equal to (a) $5$ 
and (b)  $40$.
The significance of the red(dark) regions and 
yellow(light) lines is the same as given by in Fig.~\ref{fig:theta=15}.
}
\label{fig:theta=80}
\end{figure}
Figure~\ref{fig:theta=80} shows snapshots of structures  with $\theta =80^\circ$.
In Fig.~\ref{fig:theta=80}(a),
\color{black}
a chain-like structure, in which square tetramers such as B  are connected by two bonds such as A
has formed because of a large $\chi$ value.
Hereafter, we  refer to this  chain-like structure  as chain(II). 
Another chain--like structure,  
in which triangular trimers or rhomboidal clusters are connected by single bonds, 
has formed  
under short-range attraction.
This chain-like structure is   different from the chain(II) structure
because the unit of the chain is a square tetramer in the chain(II).
We refer to this chain-like structure as chain(I).
When the pressure increases, straight chains of  chain(II)   are broken 
and bent  at the weakly connected parts 
consisting of chain(I)
[Fig.~\ref{fig:theta=80}(b)].
\color{black} 

\subsection{Structures for $\theta > 90^\circ$}

\begin{figure}[htp]
\centering
\includegraphics[width=7.0cm,clip]{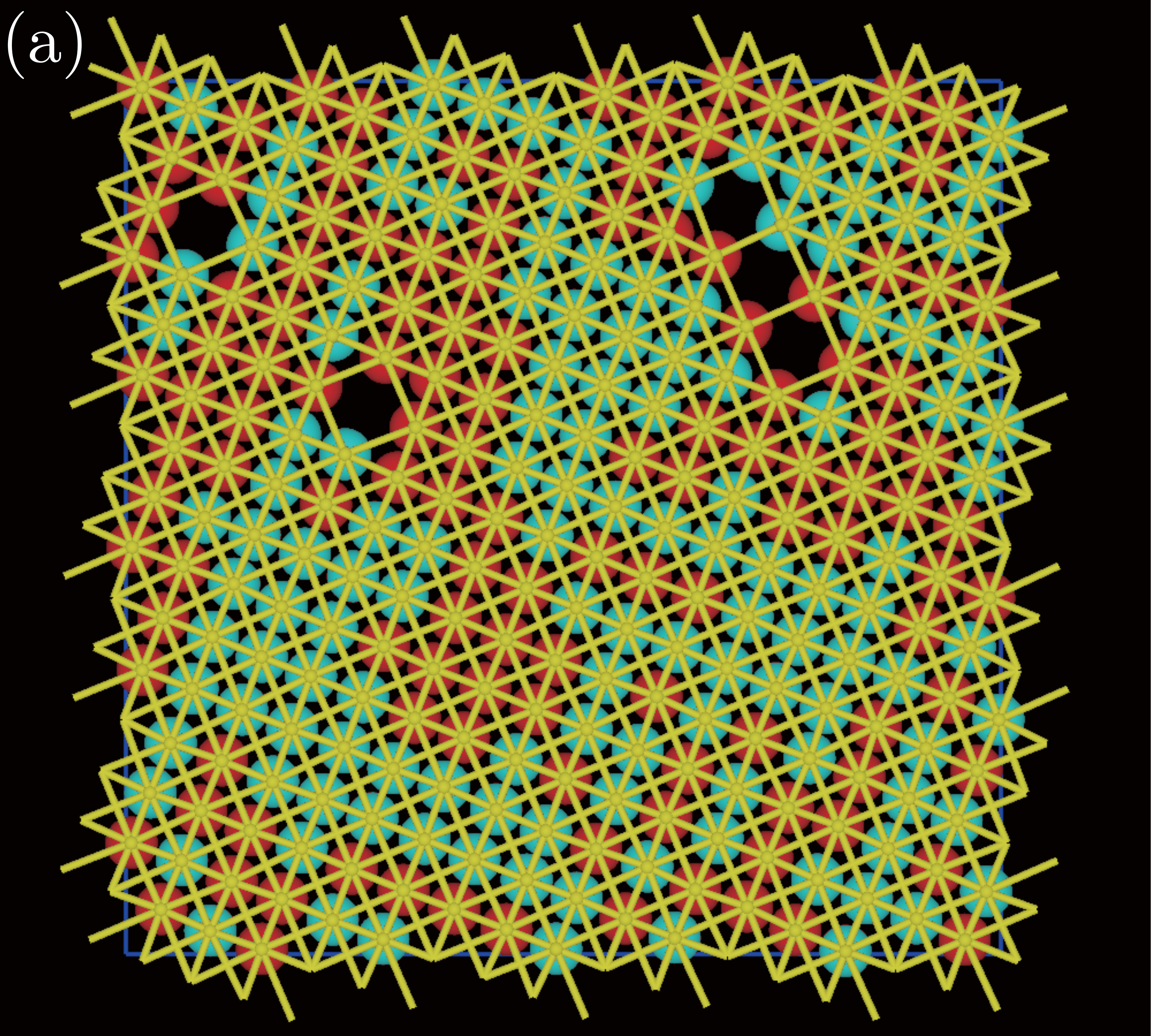} 
\includegraphics[width=7.0cm,clip]{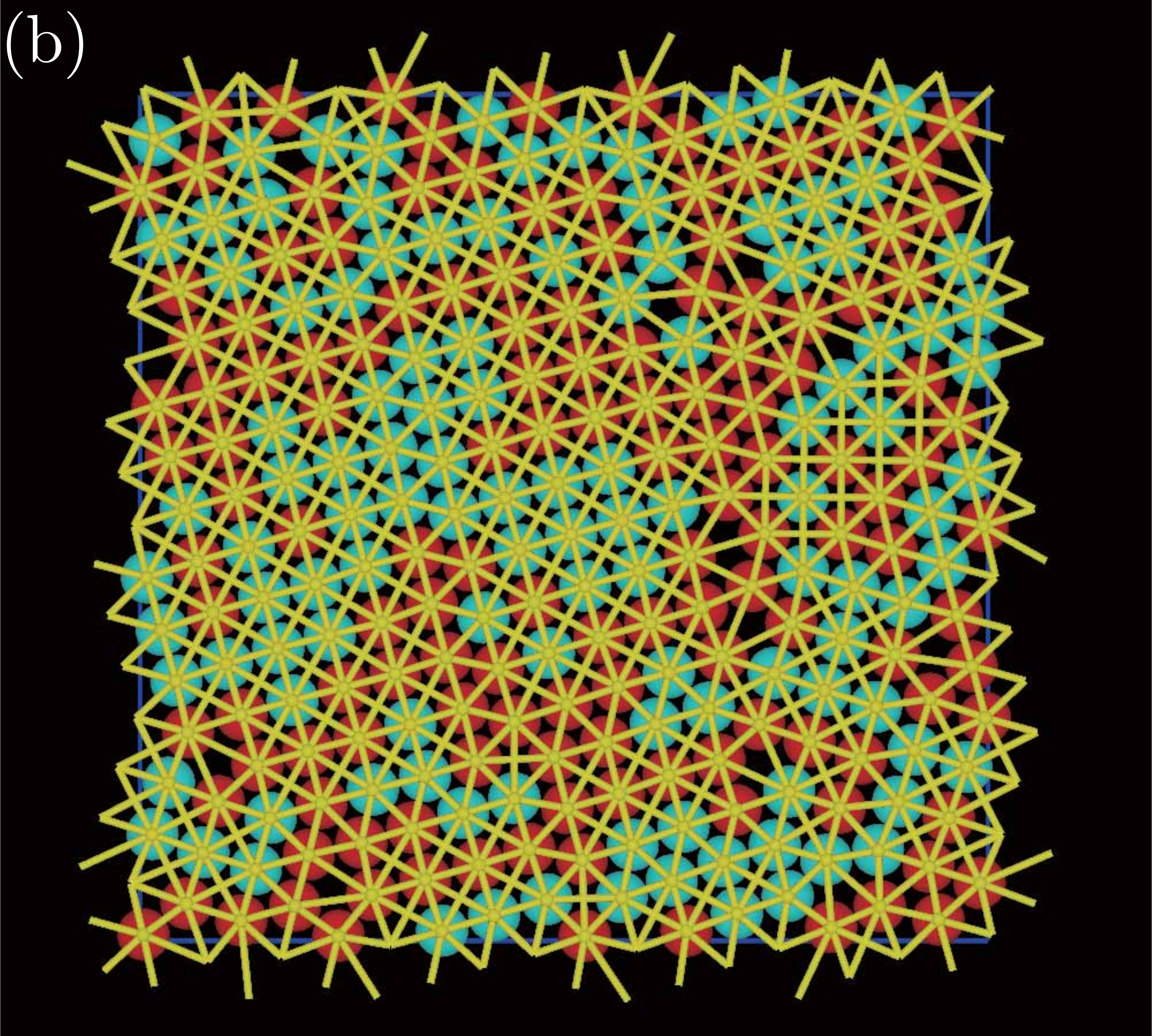}

\caption{
(color online)
Snapshots of two-dimensional structures  with $\theta =95^\circ$
and $P\sigma^3/k_\mathrm{B}T$ equal to (a) $15$ 
and (b)  $55$.
The significance of the red(dark) regions and 
yellow(light) lines is the same as given by in Fig.~\ref{fig:theta=15}.
}
\label{fig:theta=95}
\end{figure}
The chain-like clusters  shown in Fig.~\ref{fig:theta=80}  are  organized when  $\theta  \le 90^\circ$.
The structure changes into a square lattice when $\theta > 90^\circ$. 
Hereafter, we refer to the plane in which the patchy particles are placed as the $xy$ plane
and the direction perpendicular to the $xy$ plane as the $z$-direction.
Figure~\ref{fig:theta=95} shows snapshots for $\theta =95^\circ$,
in which the direction of the patch region $\hat{\bm{n}}$  is markedly 
different from $\hat{\bm{n}}$ in Figs.~\ref{fig:theta=15}--\ref{fig:theta=80}:
the component of $\hat{\bm{n}}$  is predominantly  parallel to the $xy$-plane
when $\theta \le  90^\circ$,
but $\hat{\bm{n}}$ becomes parallel or antiparallel to the $z$-direction 
when $\theta >  90^\circ$.  
The number of connected bonds is no more than five in a square lattice when $\hat{\bm{n}}$ is in the $xy$-plane.
However,  if $\hat{\bm{n}}$ is in the $z$-direction, 
the particles can interact with eight particles at most.
Hence, $\hat{\bm{n}}$ prefers to become perpendicular to the $xy$-plane to increase the number of interacting particles.

\begin{figure}[htp]
\centering

\includegraphics[width=10.0cm,clip]{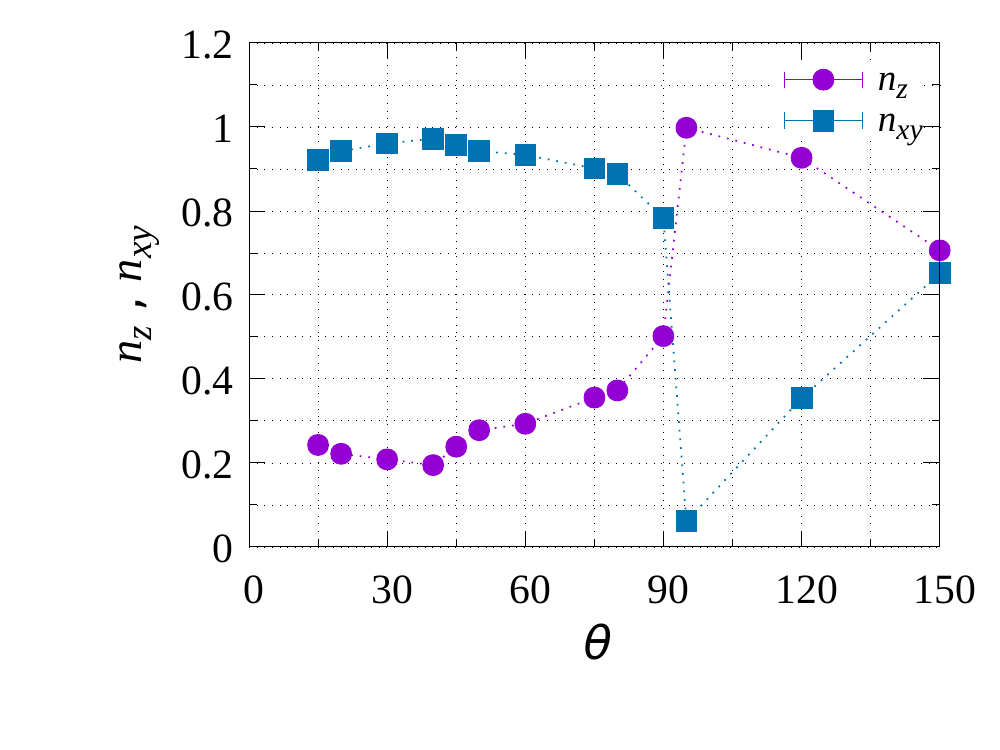} 

\caption{
(color online)
Dependence of  the amplitude of the component of  $\hat{\bm{n}}$ 
in the $z$-direction $n_\mathrm{z}$ and that parallel to the $xy$-plane $n_\mathrm{xy}$
 on the angle $\theta$, which is related to the coverage of the patch area  as $\chi=(1-\cos \theta)/2$.
The dependence is measured for pressure $P\sigma^3/k_\mathrm{B}T = 50$
and the data are averaged over 100 points \color{black} every $4 \times 10^5$ MC steps  in a run. \color{black}.
}
\label{fig:direction of patch}
\end{figure}
Figure~\ref{fig:direction of patch} shows the dependence of  the 
\color{black} average of the absolute value of the component of  $\hat{\bm{n}}$ 
in the $z$-direction $p_{z}$ and that parallel to the $xy$-plane $p_{xy}$.
They are given by 
\begin{align} 
p_z & = \left \langle  \sum_{i} |n_{i,z}| \right \rangle, \\
p_{xy} & = \left \langle \sum_{i} \sqrt{n_{i,x}^2+ n_{i,y}^2}  \right \rangle , 
\end{align}
where $\langle \cdots \rangle $ represents averaging of data in a run with a long interval of MC steps in the late stage.
\color{black}
When $\theta$ is very small, the attractive interaction between particles 
barely occurs.
Therefore,  the direction of the patch area is almost random.
The frequency of the formation of dimers increases with increasing $\theta$.
Because the directions  of the patch areas in dimers should be in the $xy$-plane, 
$n_z$  decreases with increasing $\theta$  when $\theta < 30^\circ$.
When $30^\circ<\theta<90^\circ$, the direction of the patch area fluctuates 
but maintains an  attractive interaction 
because the patch area is large.
Thus, $n_z$ increases gradually with increasing $\theta$.
When $\theta $ exceeds $90^\circ$, $n_z$ increases sharply to  increase the number of attracting particles.
However, $n_z$ decreases with increasing $\theta$ when $\theta $ is larger than $90^\circ$.
The reason is the same as that for decreasing $n_{xy}$ for  $30^\circ<\theta<90^\circ$;
that  is,
$\hat{\bm{n}}$  fluctuates but retains  the attractive interaction between particles because of the large patch area.
The lattice structure probably changes  to a hexagonal lattice for short-range attraction.
Taking into account that the change in $\hat{\bm{n}}$ at $\theta=90^\circ$ cause 
the number of attracting particles to increase,
we believe that the sharp change is also  expected
in the hexagonal lattice,
which has not been pointed out in  previous studies~\cite{Shin-Schweizer_softmatter10_2014_262,Iwashita-k_softmatter10_2014_7170}.

\color{black}
\subsection{$P\sigma^3/k_\mathrm{B}T$-- $\theta$ phase diagrams for some $\epsilon/k_\mathrm{B}T$}

\begin{figure}[htp]
\centering

\includegraphics[width=9.50cm,clip]{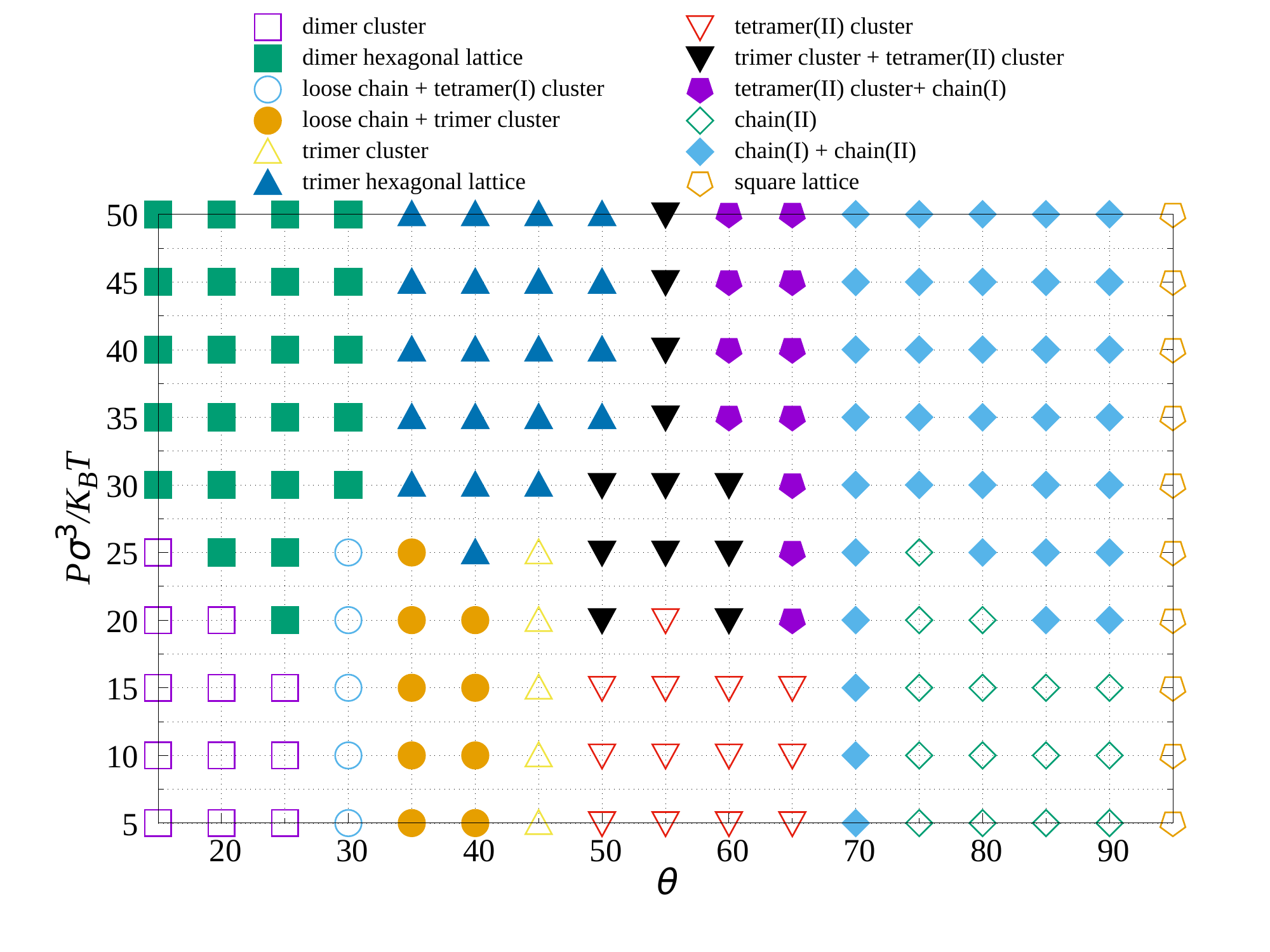} 

\caption{
(color online)
$P\sigma^3/k_\mathrm{B}T-\theta$ phase diagram for
$\epsilon/k_\mathrm{B}T=8$
}
\label{fig:phase-diagram_e8}
\end{figure}
Here, we show $P\sigma^3/k_\mathrm{B}T$--$\theta$ phase diagrams for some values.
Figure~\ref{fig:phase-diagram_e8} shows the phase diagram for $\epsilon/k_\mathrm{B}T=8$.
Here we refer  to the square tetramer formed with small $\theta $ as tetramer (I) 
and that formed with large $\theta$ as tetramer (II). 
Hexagonal lattices  form when $\theta$ is small and the pressure is high.
To judge whether hexagonal lattices form, we use a parameter $\phi_6$ which shows the local six-fold rotational symmetry.
The parameter $\phi_6$ is defined as 
\begin{equation}
\phi_6 =  \frac{1}{N}  \sum_{i}^{N} \left| \sum_{r_{ij}< \sigma^\prime} \frac{1}{6} \exp(6i\theta_{ij}) \right|,
\end{equation}
where $\theta_{ij}$ shows the angle between $\bm{r}_{ij}$ and $x$-axis. 
The summation $\sum_{r_{ij}< \sigma^\prime}$ is performed for the $j$th  particle  when $r_{ij} < \sigma^\prime$.
We set $\sigma^\prime $ to $1.2\sigma$
because the distance between the nearest neighbors is small when the pressure is high. 
$\phi_6$ should be $1$ if the perfect hexagonal lattice forms. 
However, taking account of thermal fluctuations, we consider that the hexagonal lattice forms when $\phi_6 >0.7$.
When $50^\circ \le \theta \le 60^\circ$,
the boundary between the phase with tetramer (II) clusters and that with the mixture of tetramer(II) and triangular trimer 
are determined by the ratio of $n_3$ to $n_4$ and snapshots.

\begin{figure}[htp]
\centering

\includegraphics[width=7.0cm,clip]{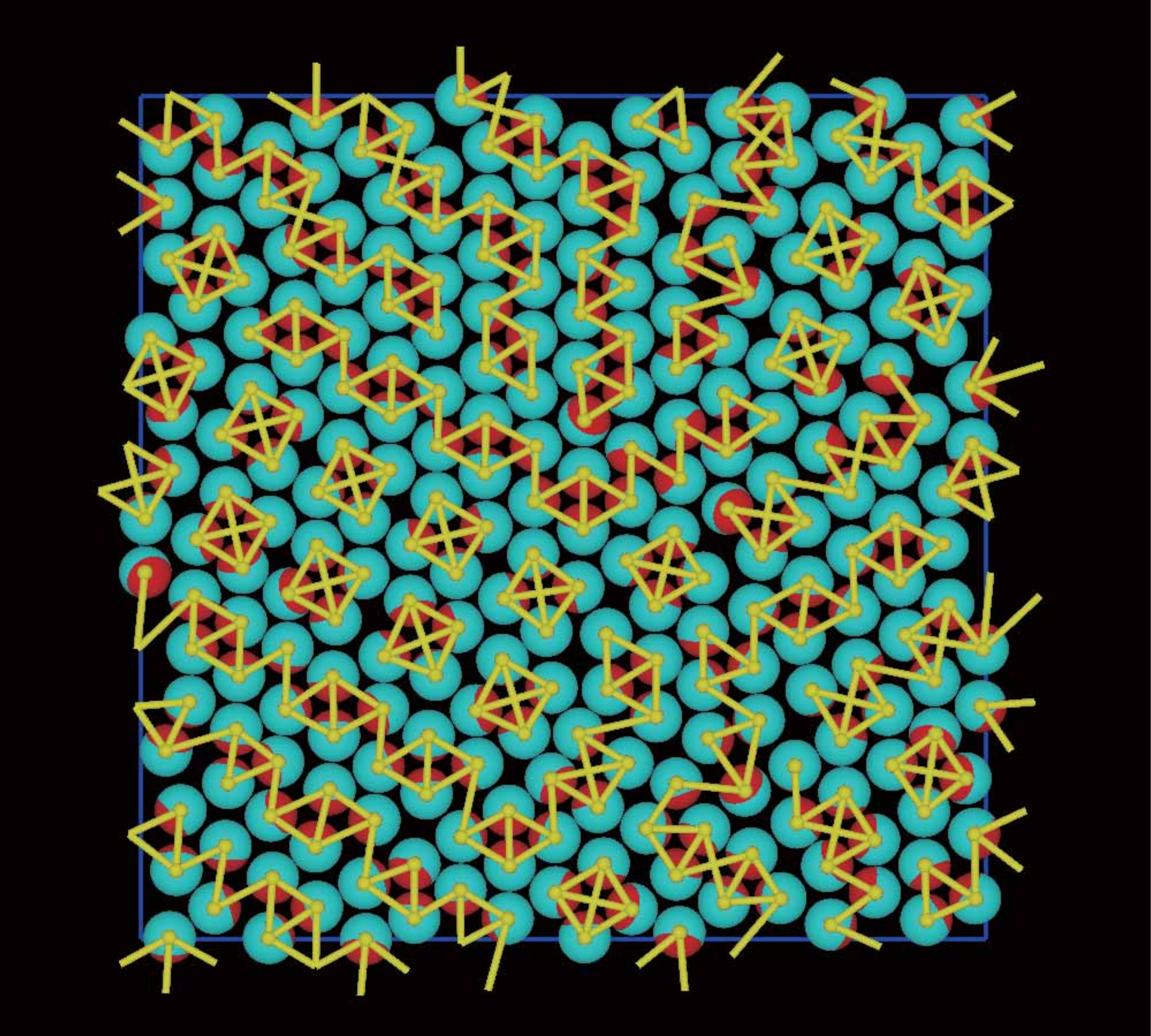} 

\caption{
(color online)
Snapshots of two-dimensional structures  with $\theta =65^\circ$
and $P\sigma^3/k_\mathrm{B}T=30$.
The significance of the red(dark) regions and 
yellow(light) lines is the same as given by in Fig.~\ref{fig:theta=15}.
}
\label{fig:theta=65}
\end{figure}
We have not shown a snapshot for the phase consisting of  tetramer(II) and chain(I). 
A typical snapshot of this phase is given by Fig.~\ref{fig:theta=65}.
The number of square tetramer(II) decreases with increasing the pressure.
When $ 70^\circ \le \theta \le 90^\circ$,
the  system transfers  from the phase of chain(II)   to that formed by  both
chain(I) and chain(II)  
with increasing the pressure.  
For simplicity,
we determined the phase  from snapshots. 
When $\theta$ changes  from $90^\circ$ to $95^\circ$, 
the direction of patch area and the structures formed by particles drastically changes;
the direction of patch area is parallel or antiparallel to the $z$-axis and 
particles form a square lattice.

\begin{figure}[htp]
\centering

\includegraphics[width=9.50cm,clip]{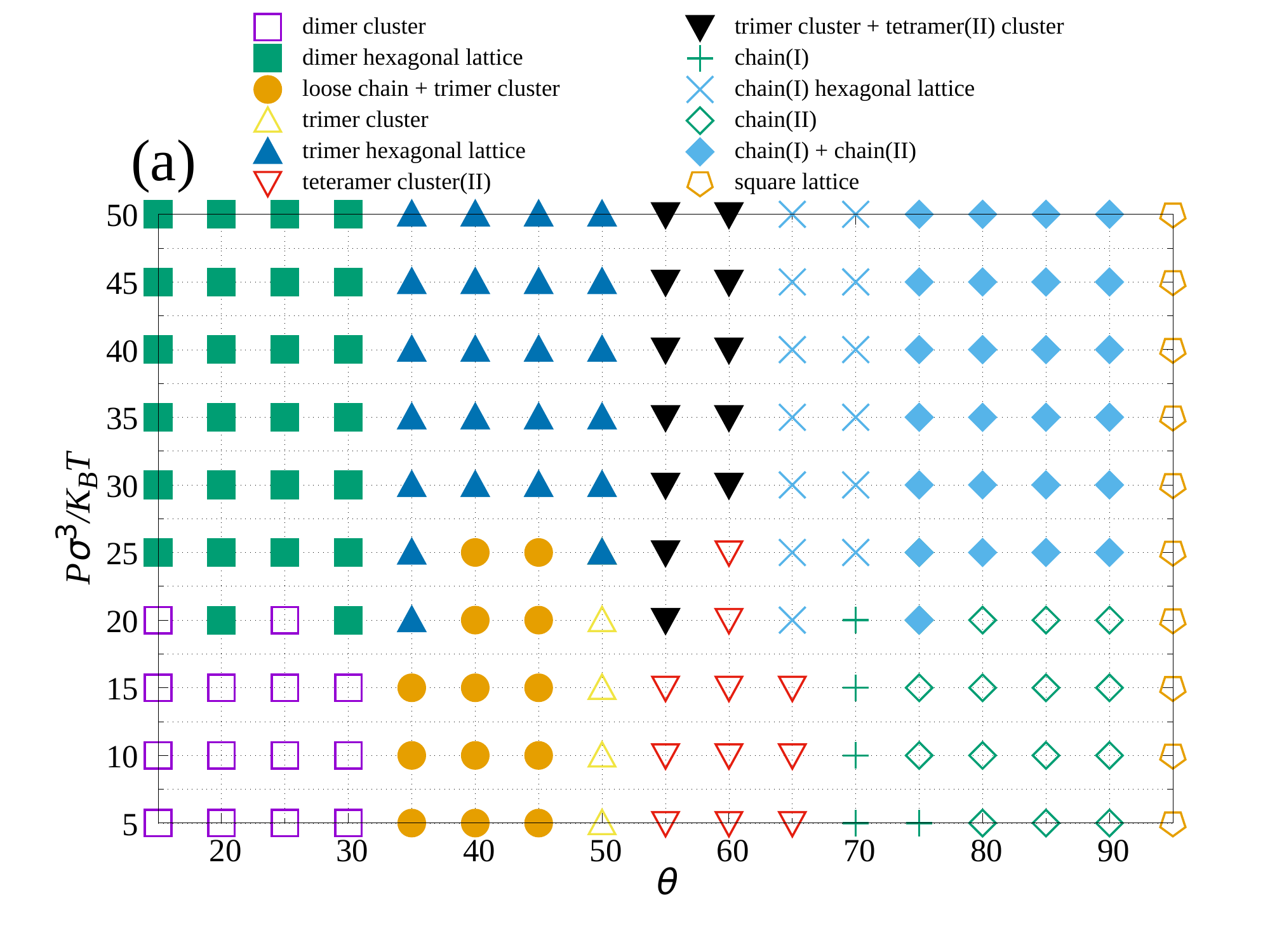} 
\includegraphics[width=9.50cm,clip]{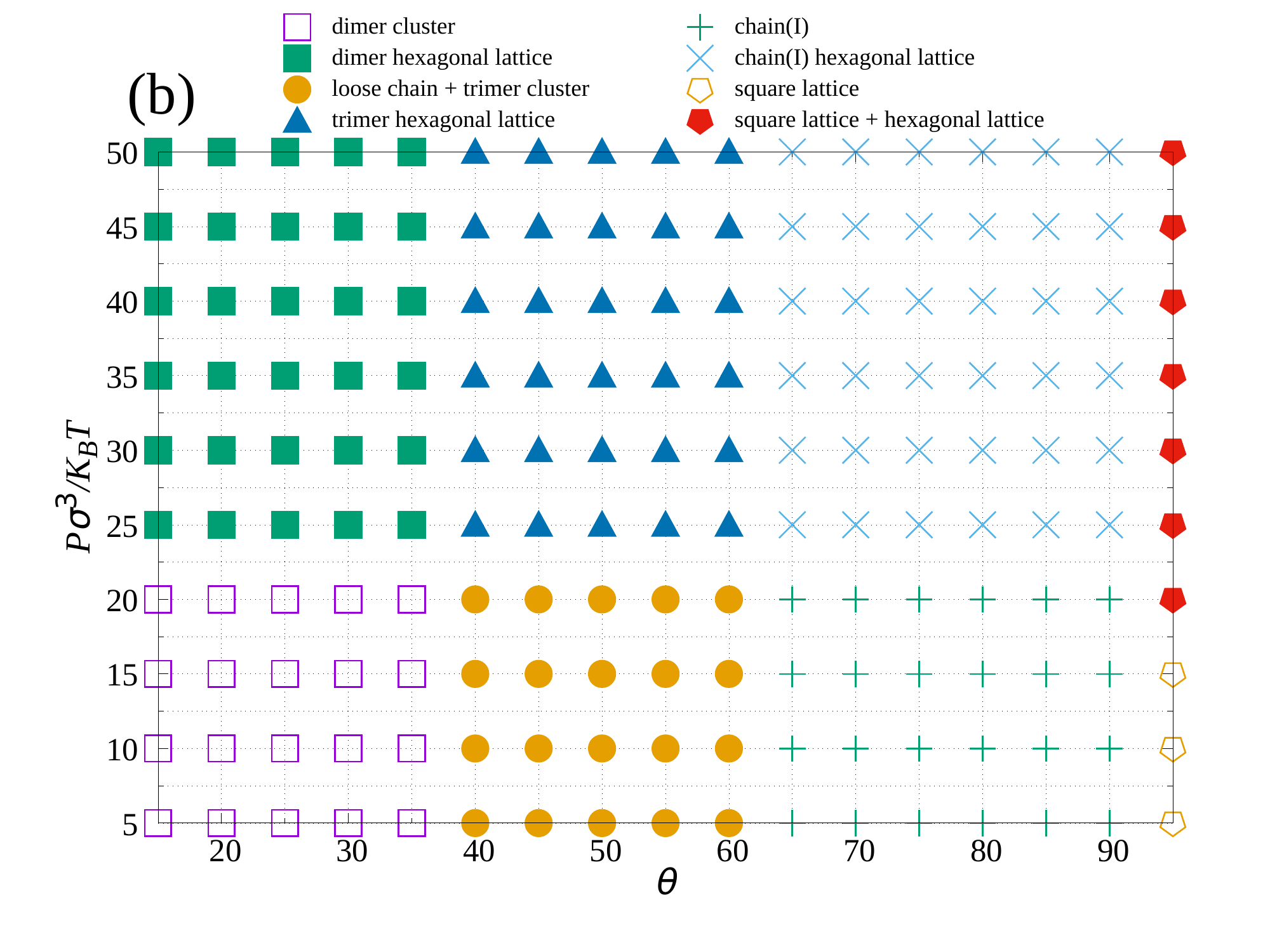} 
\includegraphics[width=9.50cm,clip]{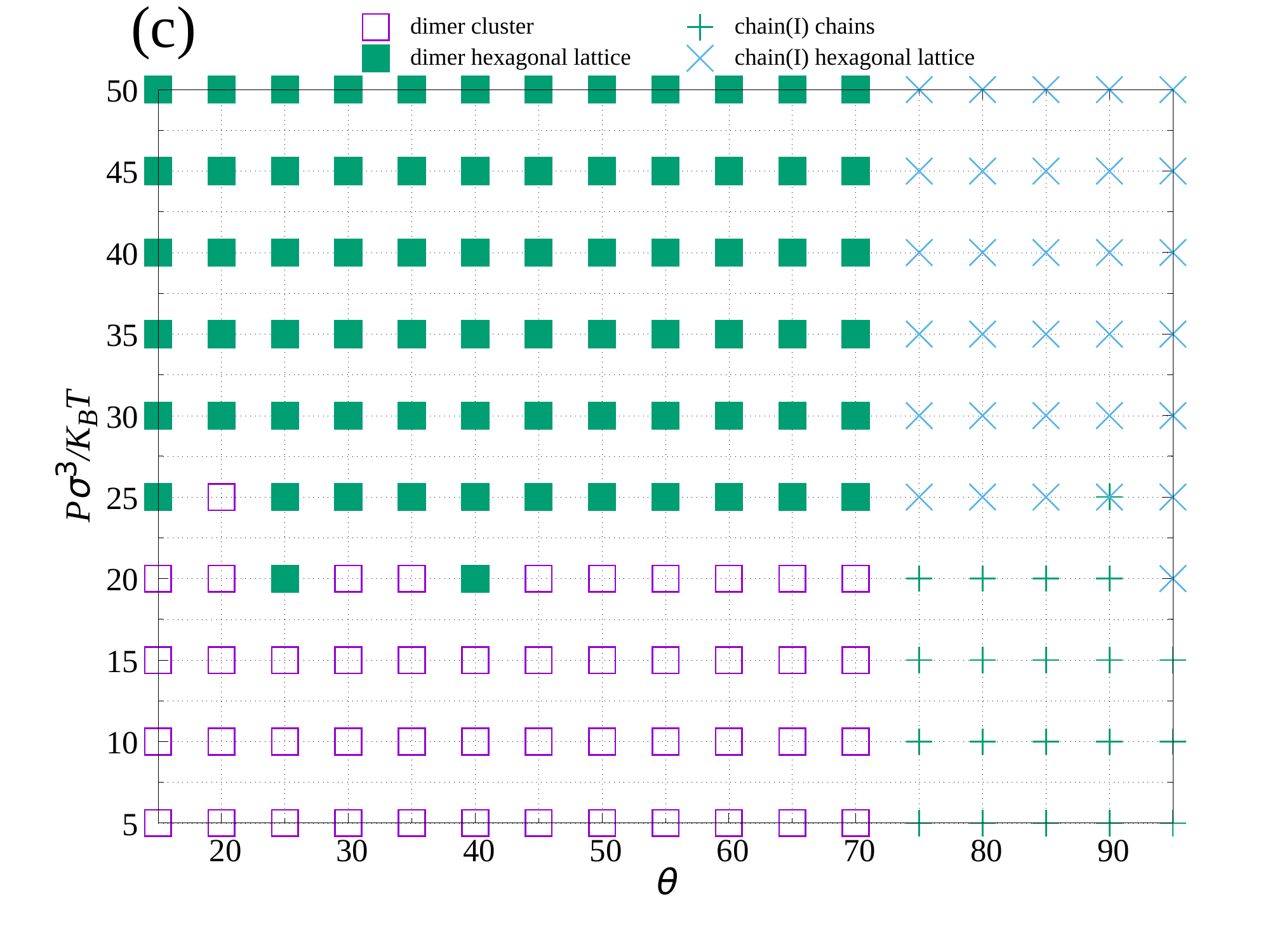} 

\caption{
(color online)
$P\sigma^3/k_\mathrm{B}T-\theta$ phase diagrams for
$\epsilon/k_\mathrm{B}T=$ (a) $6$, (b) $4$, and (c) $1$. 
}
\label{fig:phase-diagram_e6-1}
\end{figure}
\color{black}
Figure~\ref{fig:phase-diagram_e6-1} shows how the phase diagram changes with decreasing $\epsilon/k_\mathrm{B}T$.
When $\epsilon/k_\mathrm{B}T=6$ [Fig.~\ref{fig:phase-diagram_e6-1}(a)],
square tetramers and chain(II) form when the pressure is low and $\theta $ is large,
but the areas in which those structures form are slightly smaller than those in Fig.~\ref{fig:phase-diagram_e8}.
When $\epsilon/k_\mathrm{B}T=4$ [Fig.~\ref{fig:phase-diagram_e6-1}(b)],
the structures which are caused by the square tetramers disappear;
the clusters of the square tetramer, the chain(II),  and the mixture of chain (I) and chain (II) do not form.
When $\theta=95^\circ$, 
the region with a hexagonal lattice appears under high pressure
although the patch direction in the region is parallel or antiparallel to the $z$-axis.
When $\epsilon/k_\mathrm{B}T=4$ [Fig.~\ref{fig:phase-diagram_e6-1}(c)],
triangular trimers and the hexagonal lattice formed by them do not form.
Dimers and hexagonal lattice formed by dimers appear instead.
In the regions where these structures form,
triangular trimers and short zigzag chain appear with increasing  $\theta$,
but the their numbers are very small.

\subsection{Effect of the form of $U_\mathrm{att}$on structures}
In our simulations, a main difference from previous 
studies~\cite{Shin-Schweizer_softmatter10_2014_262,Iwashita-k_softmatter10_2014_7170}
 is the formation of square tetramers. 
Since $\Delta $ is set to $\sigma /2$, the particles in the diagonal positions can 
attract each other  in  the square cluster whose side length is $\sigma$.
We think that  the square tetramers  do not form when  $\Delta \le (\sqrt{2}-1) \sigma$
because the attraction between the particles in the diagonal positions in the square 
tetramers vanishes.

\begin{figure}[htp]
\centering
\includegraphics[width=7.0cm,clip]{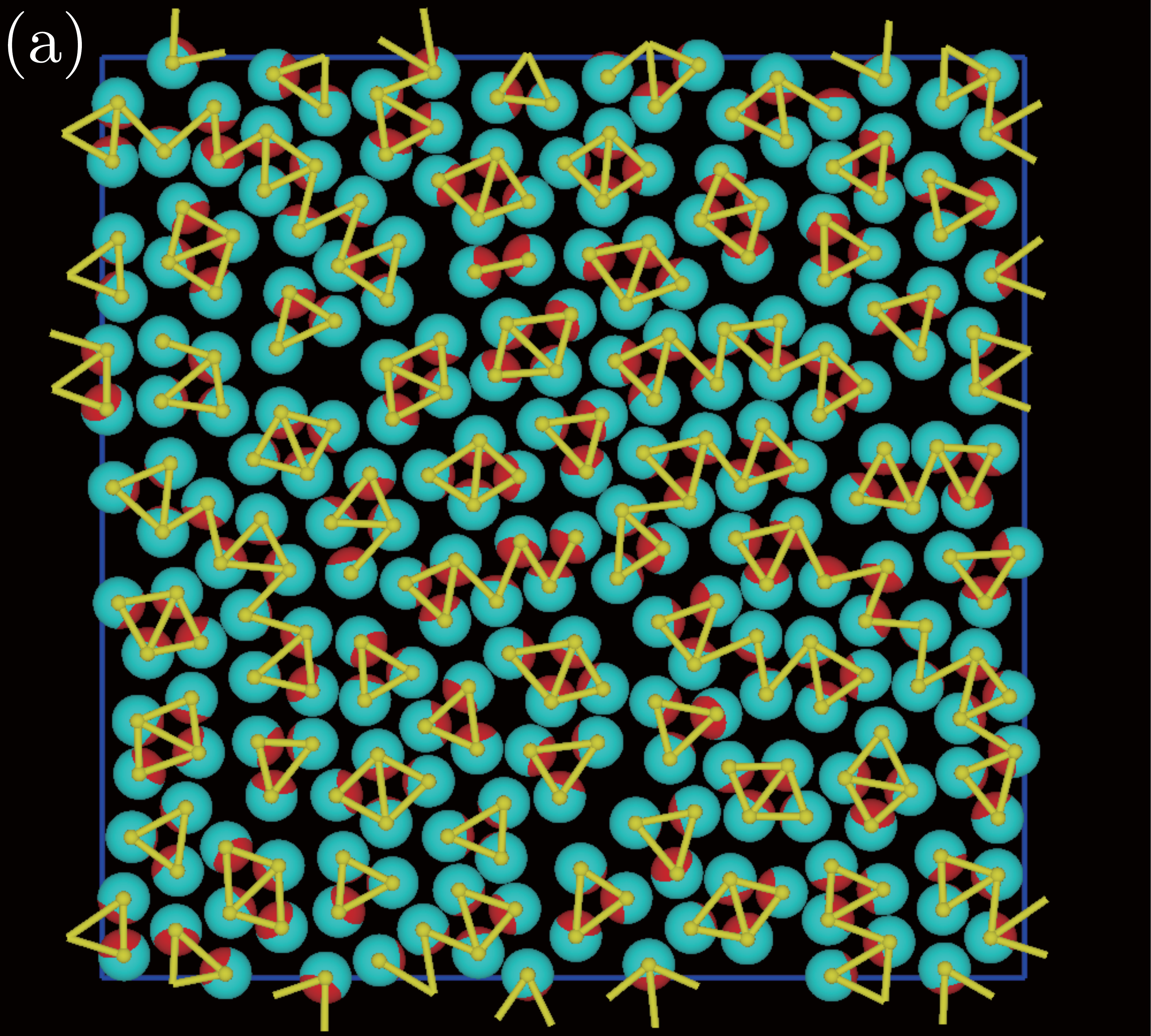} 
\includegraphics[width=7.0cm,clip]{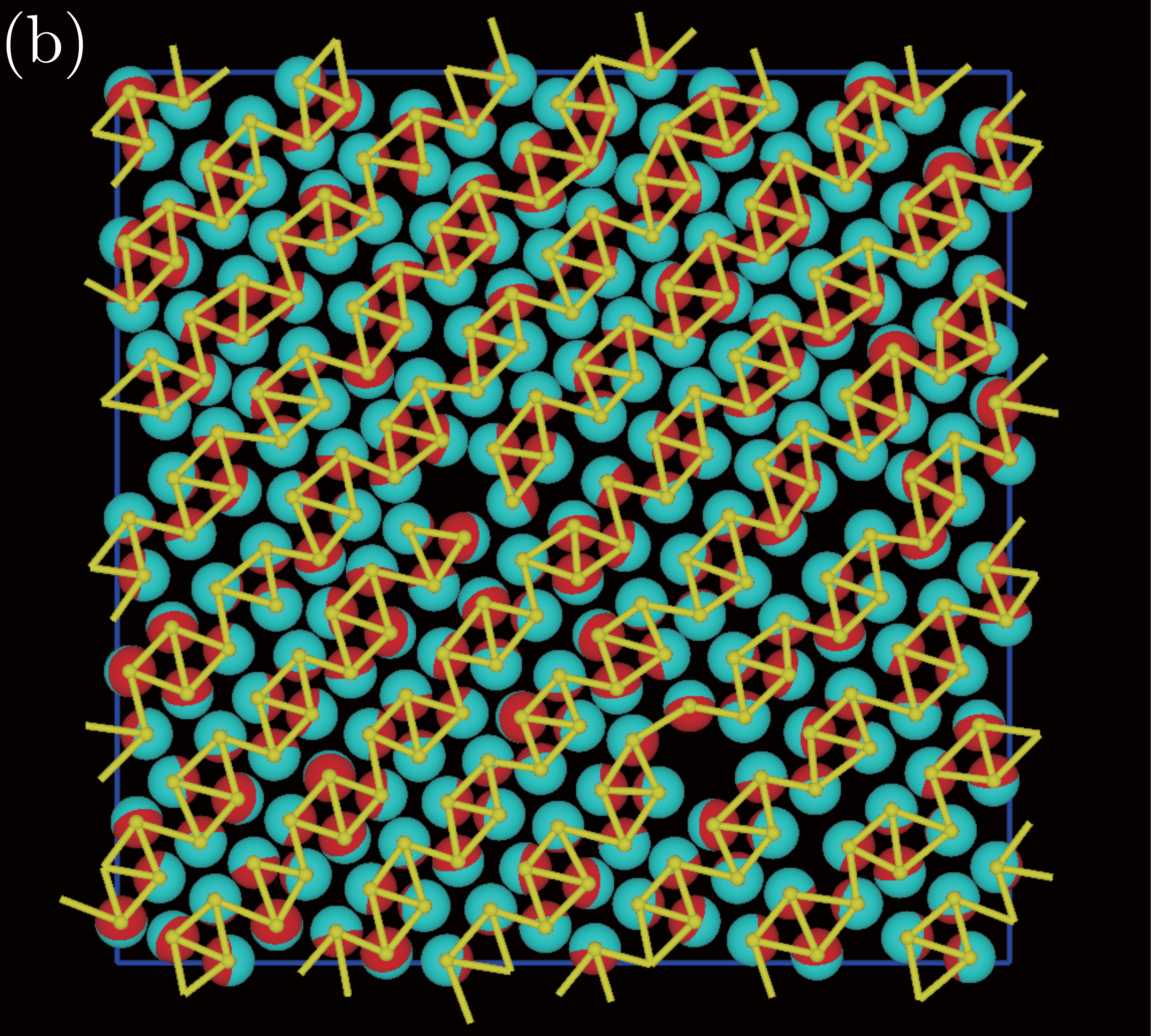}
\caption{ 
(color online)
Snapshots of two-dimensional structures 
for $\theta =$ (a) $60^\circ$ and (b) $80^\circ$.
$P\sigma^3/k_\mathrm{B}T$  equal to $5$
and $\Delta$ is set to $0.3$.
The significance of the red(dark) regions and 
yellow(light) lines is the same as given by in Fig.~\ref{fig:theta=15}.
}
\label{fig:theta=60and80delta1.5}
\end{figure}
To confirm the prediction, we set $\Delta $ to $0.3\sigma <   (\sqrt{2}-1) \sigma  $ and performed simulations.
Figures~\ref{fig:theta=60and80delta1.5}(a) and (b)  show snapshots for 
$\theta=60^\circ$
and $\theta=80^\circ$.
Square tetramers observed in Fig.~\ref{fig:theta=50} do not form 
in Fig.~\ref{fig:theta=60and80delta1.5}(a), and triangular trimers and rhomboidal clusters
form instead [Fig.~\ref{fig:theta=60and80delta1.5}(a)].
The form of chains also changes from chain (II)
 to chain (I) when $\theta=80^\circ$ [Fig.~\ref{fig:theta=60and80delta1.5}(b)].
\begin{figure}[htp]
\centering
\includegraphics[width=7.0cm,clip]{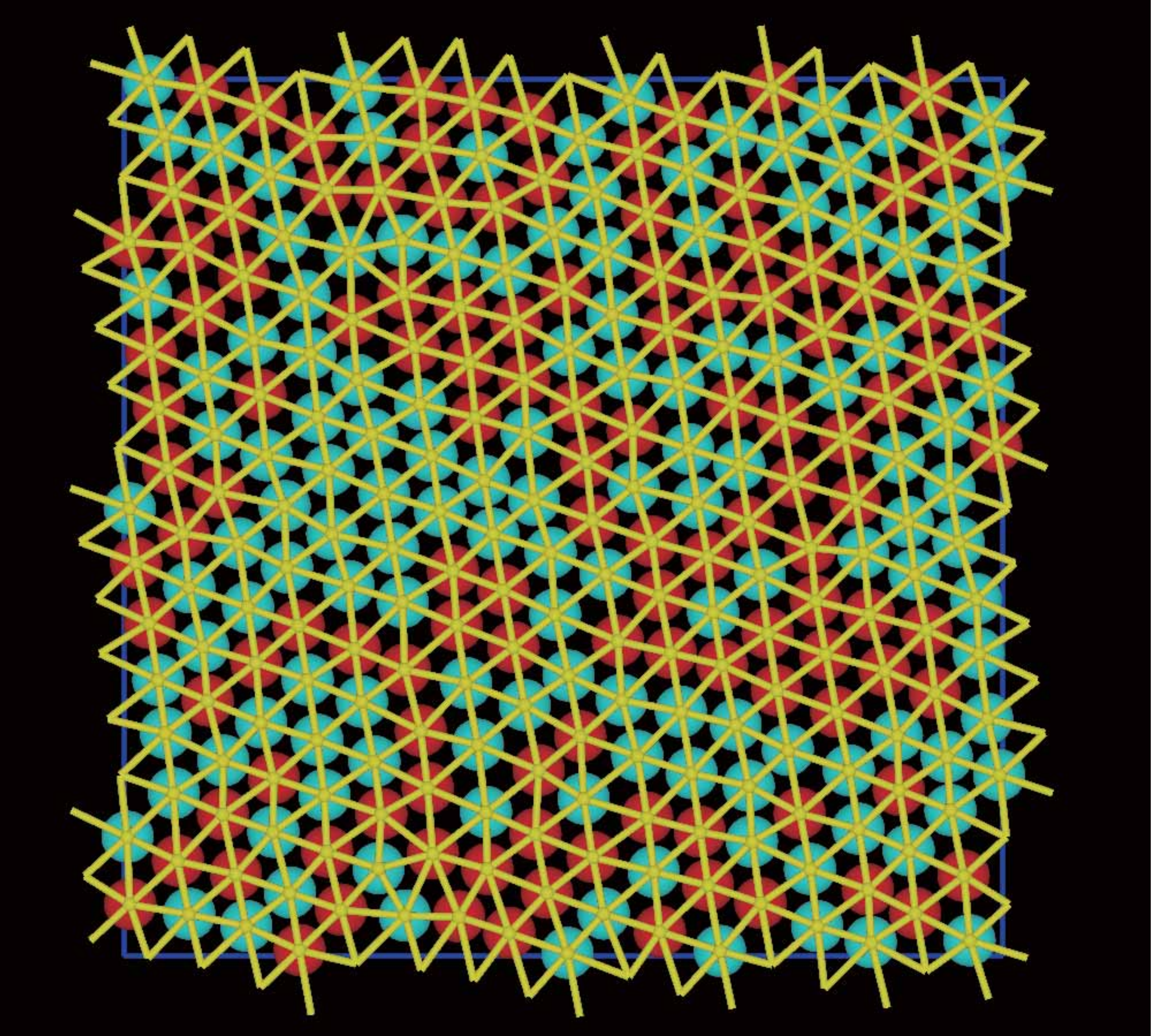} 

\caption{
(color online)
A snapshot of two-dimensional structure  with $\theta =95^\circ$,
$\Delta =0.3 \sigma <  (\sqrt{2}-1) \sigma$, and $P\sigma^3/k_\mathrm{B}T=5$.
The significance of the red(dark) regions and 
yellow(light) lines is the same as given by in Fig.~\ref{fig:theta=15}.
}
\label{fig:theta=95delta1.5}
\end{figure}
The structure for $\theta >90^\circ$ also changes.
The direction of patch area for each particle is perpendicular to the plane where the particles are located,
but the square lattice observed in Fig.~\ref{fig:theta=95} changes to the hexagonal lattice as shown 
in Fig.~\ref{fig:theta=95delta1.5}.
These results agree with our prediction.

We also performed simulations using the Lennard-Jones (LJ) potential as $U_\mathrm{att}(r)$.
We considered the hard-core repulsive potential when $r_\mathrm{ij}<2^{1/6}\sigma$.
$U_\mathrm{att}(r_{ij}) $ is given by 
\begin{equation}
U_\mathrm{att}(r_{ij})
=
4\epsilon \left \{ \left( \frac{\sigma}{r} \right)^{12} - \left( \frac{\sigma}{r} \right)^{6} \right \}
    \quad  (2^{1/6} \sigma < r_{ij} ).\\
\label{eq:LJ}
\end{equation}
We set $\epsilon/k_\mathrm{B}T =8$ in the simulations
for the minimum of the attractive potential to be $-\epsilon$.
\begin{figure}[htp]
\centering
\includegraphics[width=7.0cm,clip]{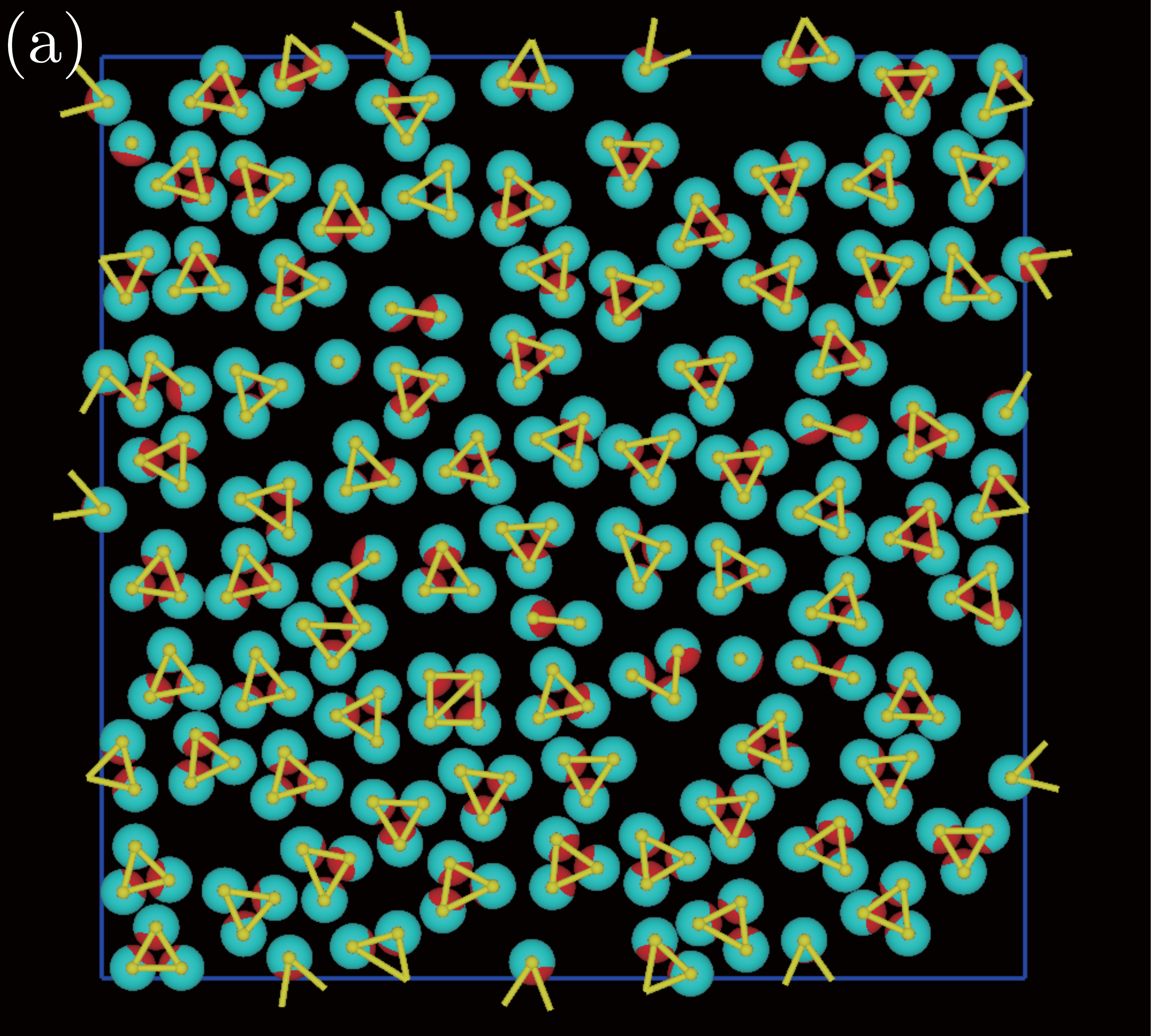} 
\includegraphics[width=7.0cm,clip]{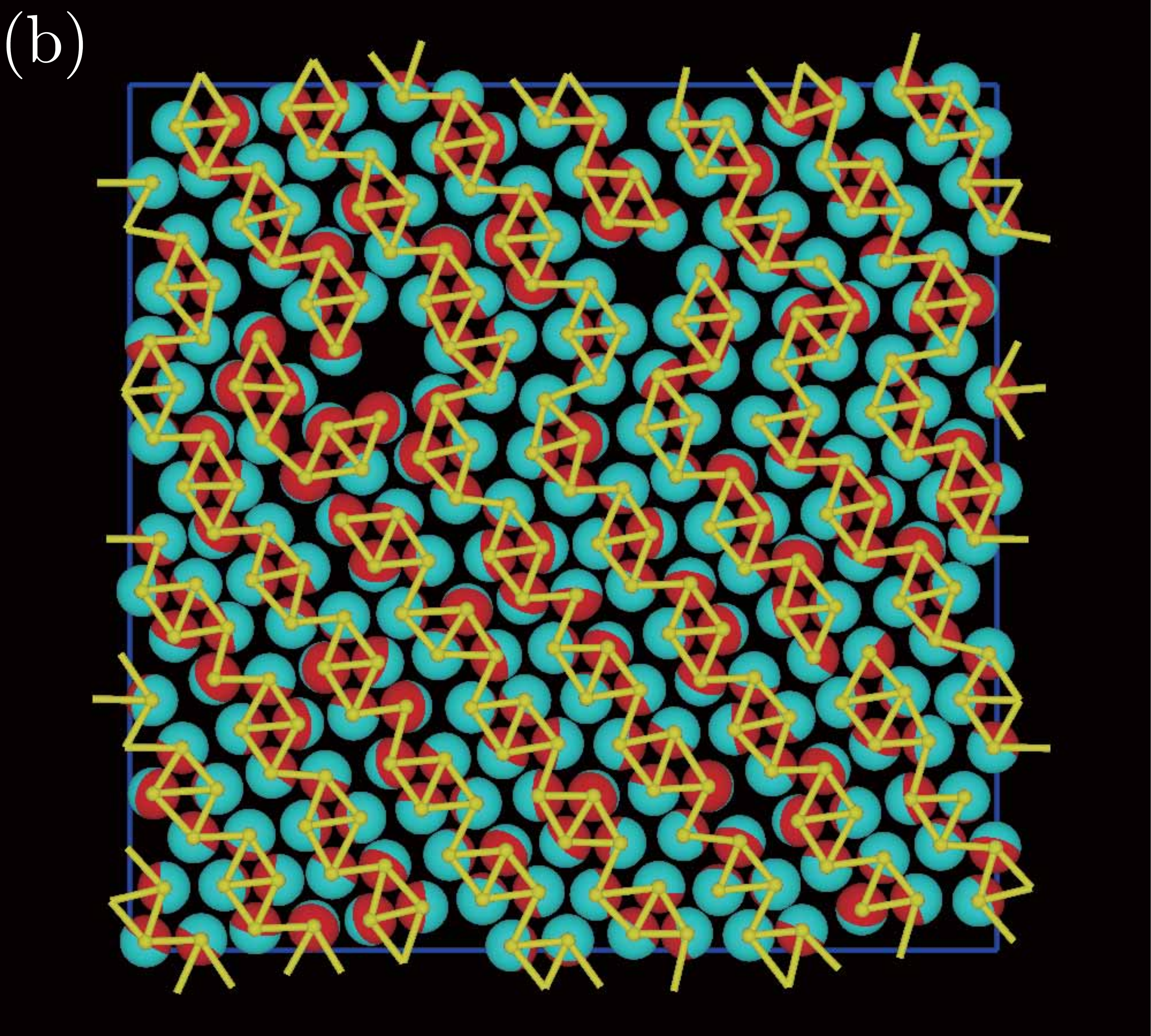}
\caption{ 
(color online)
Snapshots of two-dimensional structures for the LJ attractive potential.
$\theta $ and $P\sigma^3/k_\mathrm{B}T$  
are (a) $55^\circ$  and $5$ and (b) $85^\circ$ and $10$, respectively.
The significance of the red(dark) regions is the  as given by in Fig.~\ref{fig:theta=15}.
yellow(light) lines are drawn when the distance between the centers of particles is smaller than $2^{1/6} \sigma$.
}
\label{fig:theta=55and85LJ}
\end{figure}
Figures~\ref{fig:theta=55and85LJ}(a) and (b)  show snapshots for $\theta=55^\circ$ and $85^\circ$, respectively.
When $U_\mathrm{att}(r)$ is the square wall potential with $\Delta =\sigma/2$,
the square tetramers and chain (II)  form for these $\theta$.
However, when the attractive potential is the LJ potential,
triangular trimers and single chains form instead to increase the number of the nearest neighbors.

\begin{figure}[htp]
\centering
\includegraphics[width=7.0cm,clip]{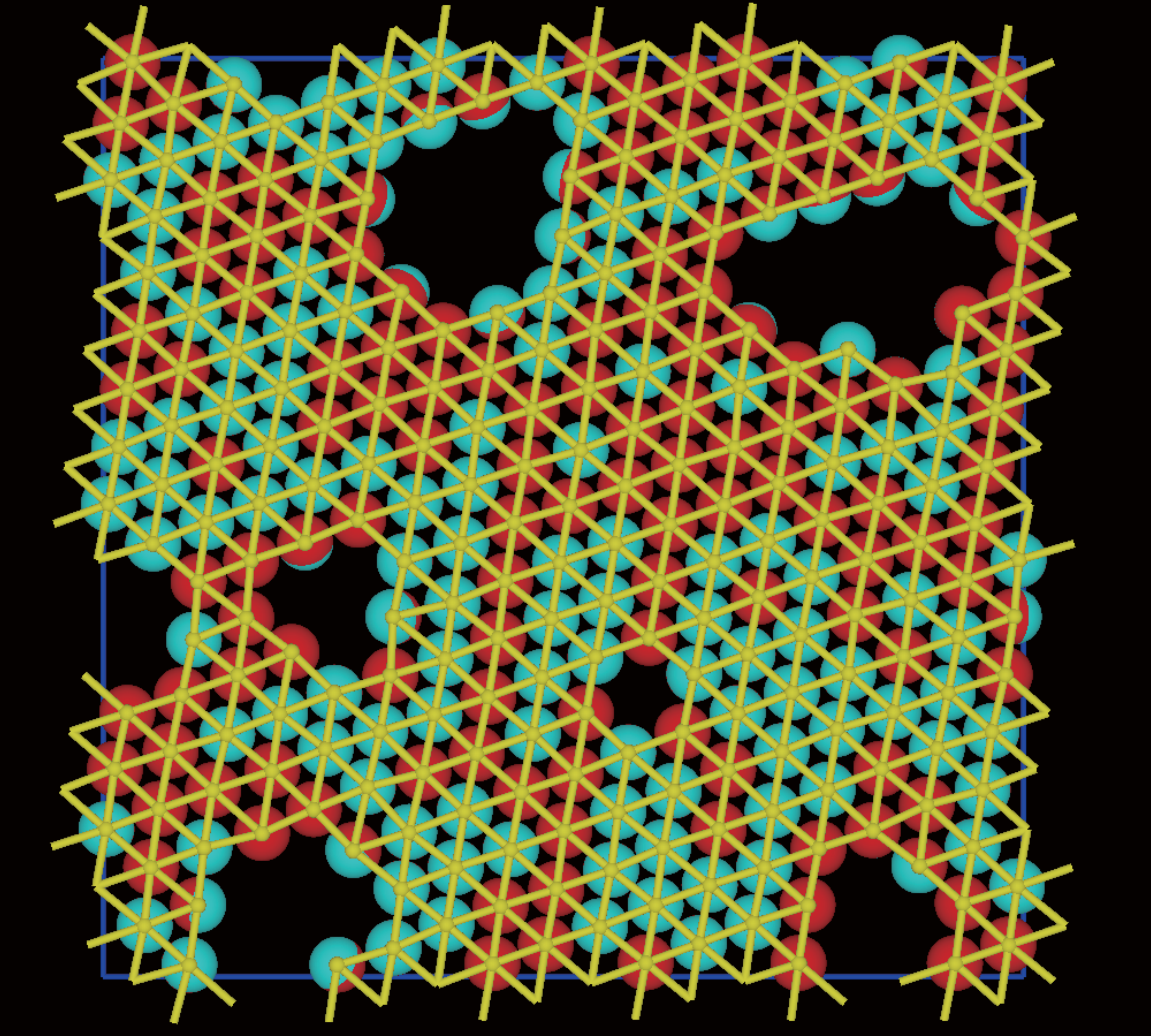} 

\caption{
(color online)
Snapshots of two-dimensional structures for the LJ attractive potential.
$\theta =95^\circ$ and $P\sigma^3/k_\mathrm{B}T=30$.
The significance of the red(dark) regions
yellow(light) lines is 
the same as given by Fig.~\ref{fig:theta=55and85LJ}
}
\label{fig:theta=95LJ}
\end{figure}
Figure~\ref{fig:theta=95LJ} shows a  snapshot for $\theta=95^\circ$.
Although large voids remain even with the high pressure, 
the hexagonal lattice forms for the LJ potential as well as for the square wall potential with  $\Delta < \sigma/2$
because 
the number of the nearest neighbors is the largest in the hexagonal lattice.
\color{black}

\section{Summary}\label{sec:summary}
We performed both Monte Carlo simulations and Brownian dynamics simulations
to study self-assemblies formed by one-patch particles.
Using  the KF potential in Monte Carlo simulations,
we studied how  the self--assemblies with long-range attractive interactions differ from those with 
short-range attractive interactions.
The clusters formed in low pressure changed from dimers to triangle trimers with increasing $\chi$.
Square tetramers, \color{black} chains formed by square tetramers \color{black}, and islands with a square lattice
are also formed with further increase in $\chi$.

Although the system size of our simulations is not so large, we obtained 
new results. 
The main difference between our results and previous results from a short-range attractive interactions is 
the formation of loose zigzag chain  and square tetramers.
The loose zigzag chain is observed when $\chi$ is small.
In our simulations, two types of square tetramers are formed.
When $\chi$ is small, the particles in the square tetramers interact with one of neighbors  and the particle in the diagonal position
but do not interact with the other neighbor.
In the other, when $\chi$ is large,
the particles in the square tetramers can interact with all other particles.
These self-assemblies form under low pressure.


The other main result  is that the patch direction $\hat{\bm n}$ changes 
\color{black}  at $\theta =90^\circ$ when $\epsilon/k_\mathrm{B}T \ge 6$.
\color{black}
$\hat{\bm n}$ is in the plane where the particles are located when $\theta \le 90^\circ$, 
but changes sharply to lie perpendicular to the plane when $\theta$ exceeds $90^\circ$.
The change in $\hat{\bm n}$ is not because of a long-range attraction.
As the direction of the patch region changes to increase the number of neighbors,
the same change in $\hat{\bm n}$ should also occur for short-range attraction.
The point has  not been mentioned in the literature~\cite{Shin-Schweizer_softmatter10_2014_262,%
Iwashita-k_softmatter10_2014_7170}

In our simulations,
the formation of new types of self-assemblies arise from the
\color{black} square wall like \color{black} 
long-range attraction.
It may be difficult to realize experimentally such long-range attractions,
but we suggest that coating particles with DNA strands is one possible method to create the potential.
Currently, 
designing DNA freely and controlling the interaction between particles to produce a structure as desired
is becoming possible~\cite{Nykypanchuk,Park,Macfarlane_PNAS,Macfarlane_Science,Znag,Isogai-jcg,Isogai-jjap}.
We remain hopeful  that patchy particles with long-range attractive potentials are  realized 
through this technique.

\begin{acknowledgments}
This work was supported by JSPS KAKENHI Grants,
Nos. JP16K05470, JP18K04960, and  JP18H03839,
and the Grant for Joint Research Program of the Institute 
of Low Temperature Science, Hokkaido University, Grant number 19G020.
\end{acknowledgments}



\begin{thebibliography}{999}




\bibitem{Zhan-Nanolett4_2004_1407}
Z. Zhang and S. C. Glotzer, 
Nano Lett.
\textbf{4}, 1407 (2004)

\bibitem{Maldovan_natMater3_2004593}
M. Maldovan and E. L. Thomas, 
Nat. Mater. \textbf{3}, 593 (2004).

\bibitem{Roh_natMater4_2015_759}
K.-H. Roh, D. C. Martin, and J. Lahann,
Nat. Mater. \textbf{4}, 759 (2005).


\bibitem{Wang_Nature491_2012_51}
Y. Wang, Y. Wang, D. R. Breed, V. N. Manoharan, 
L. Feng, A. D. Hollingsworth, M. Weck, and D. J. Pine, 
Nature (London) \ textbf{491}, 51 (2012).

\bibitem{Mao-G_natMater12_2013_217}
X. Mao, Q. Chan, and S. Granick,
Nat. Mater. \textbf{12}, 217 (2013).


\bibitem{Glotzer_natMater6-557_2007}
S. C. Glotzer, and M. J. Solomon,
Nat. Mater. \textbf{6}, 557 (2007).


\bibitem{Sacanna_nature464_2010_575}
S. Sacanna, W. T. M. Irvine, P. M, Chaikin, and D. J. Pine,
Nature  (London) \textbf{464}, 575 (2010).

\bibitem{Matthew-natMater9_2010_913}
M. R. Jones, R. J. Macfarlane, B. Lee, J. Zhang, K. L. Young,
A. J. Senesi, and C. A. Mirkin,
Nat. Mater. \textbf{9}, 913 (2010).

\bibitem{Kraft-J.Phy.Chem.B115_2011_7175}
D. J. Kraft, J. Hihorst, M. A. P. Heinen, M. J. Hoogenraad, B. Luigjes, and W. K. Kegal,
J. Phys. Chem. B \textbf{115}, 7175 (2011).

\bibitem{Kraft-pnas109_2012_10787}
D. J. Kraft, R. Ni, F. Smallenburg, M. Hermes, K. Yoon, D. A. Weitz, A. v. Blaaderen,
J. Groenevold, M. Dijsra, and W. K. Kegal,
Proc. Natl. Acad. Sci. \textbf{109}, 10787 (2012). 

\bibitem{Avvisat-jcp142_2015_084905}
G. Avvisati, T. Vissers, and M. Dijkstra,
J. Chem. Phys. \textbf{142}, 084905 (2015).


\bibitem{Geuchies-natMater15_2016_1}
J. J. Geuchies, C. van. Overbeek, W. H. Evers, B. Goris, A. de. Becker,
A. P. Gantapara, F. T. Rabouw, J. Hilhorst, J. L. Peters, O. Konovalov,
A. V. Petukhov, M. Dijksra, L. D. A. Siebbeles, S. van. Aert, S. Bals, 
and D. Vanmaekelbergh
Nat. Mater. \textbf{15}, 1 (2016).

\bibitem{Wolters-Langmuir33_2017_3270}
J. R. Wolters, J. E. Verweij, G. Avvisati, M. Dijkstra, and W. K. Kegal,
Langmuir \textbf{33}, 3270 (2017).

\bibitem{Kang}
c. Kang and A. Honciuc,
ACS Nano \textbf{12}, 3741 (2018).



\bibitem{Miller-PRE_2009_021404}
W. L. Miller and A. Cacciuto,
Phys. Rev. E \textbf{80}, 021404 (2009).

\bibitem{Chen-Nature469_2011_381}
Q. Chen, S. C. Bae, and, S. Granick,
Nature \textbf{469}, 381 (2011).

\bibitem{Chan-Science331_2011_199}
Q. Chen, J. K. Whitmer, S. Jiang, S. C. Bae, E. Luijten, and S. Granick,
Scinece \textbf{331}, 199 (2011).

\bibitem{Chen-Langmuir28_2012_13555}
Q. Chen, J. Yan, J. Zhang, S. C. Bae, and S. Granick,
Langmuir \textbf{28}, 13555 (2012).

\bibitem{Romano-jpcm24_2012_064113}
F. Romano, E. Sanz, P. Tartaglia, and F. Sciortino,
J. Phys.: Condens. Matter \textbf{24}, 064113 (2012).

\bibitem{Vissers-psds-jcp138_2013_164505}
T. Vissers, Z. Preisler, F. Smallenburg, M. Dijkstra, and F. Sciortino,
J. Chem. Phys. \textbf{138}, 164505 (2013).


\bibitem{Preisler-vsms-jpcB117_2013_9540}
Z. Preisler, T. Vissers, F. Smallenburg, G. Muna\`o, and F. Sciortino,
J. Phys. Chem. B \textbf{117}, 9540 (2013).

\bibitem{Vissers-JCP140_2014_144902}
T. Vissers, f. Smallenburg, G. Muna\`o, Z. Perisler,
and F. Sciortino,
J. Chem. Phys. \textbf{140}, 144902 (2014).

\bibitem{Preisler-vmsf_solfmatter10_2014_5121}
Z. Preisler, T. Vissers, G. Muna\`o, f. Smallenburg, and F. Sciortino,
Soft Matter \textbf{10}, 5121 (2014).

\bibitem{Shin-Schweizer_softmatter10_2014_262}
H. Shin and K. S. Schweizer,
Soft Matter \textbf{10}, 262 (2014).


\bibitem{Iwashita-k_softmatter10_2014_7170}
Y. Iwashita and Y. Kimura,
Soft Matter \textbf{10}, 7170 (2014).


\bibitem{Preisler-vss-jcp145_2016_064513}
Z. Preisler, T. Vissers, F. Smallenburg, and F. Sciortino,
J. Chem. Phys. \textbf{145}, 064513 (2016).

\bibitem{Iwashita_Scientific_reprot6_2016_27599}
Y. Iwashita and Y. Kimura,
Sci. Rep. \textbf{6}, 27599 (2016).

\bibitem{Gong_nature550_2017_234}
Z. Gong, T. Hueckel, G.-R. Yi, and S. Sacanna,
Nature \textbf{550}, 234 (2017).

\bibitem{Patra-PRE96_2017_022601}
N. Patra and A. V. Tkachenko,
Phys. Rev. E \textbf{96}, 022601 (2017).


\bibitem{Blaaderen}
A. van Blaaderen, R. Ruel, and P. Wiltzius,
Nature (London) \textbf{385}, 321 (1997).



\bibitem{Kern-f-jcp118_9882_2003}
N. Kern and D. Frenkel,
J. Chem. Phys. \textbf{118}, 9882 (2003).


\bibitem{Xia_Nat.Nanotechnol6_2011_580}
Y. Xia, T. D. Nguyen, M. Yang, B. Lee, A. Santos, P. Podsiadlo,
Z. Tang, S. C. Glotzer, and N. A. Kotov,
Nat. Nanotechnol. \textbf{6}, 580 (2011).

\bibitem{Ngyyen_Proc.Natl.Acad.Sci112_2015_E3161}
T. D. Nguyen, B. A. Schultz, N. A. Kotov, and S. C. Glotzer,
Proc. Natl. Acad. Sci. \textbf{112}, E3161 (2015).








\bibitem{Nykypanchuk}
D. Nykypanchuk, M. M. Maye, D. van der Lelie, and O. Gang, 
Nature \textbf{451}, 549 (2008).

\bibitem{Park}
S. Y. Park, A. K. R. Lytton-Jean, B. Lee, S. Weigand, G. C. Schatz, 
and C. A. Mirkin, Nature \textbf{451}, 553 (2008).

\bibitem{Macfarlane_PNAS}
R. J. Macfarlane, B. Lee, H. D. Hill, A. J. Senesi, S. Seifert, 
and C. A. Mirkin, Proc. Natl. Acad. Sci. U.S.A. \textbf{106}, 10493 (2009).

\bibitem{Macfarlane_Science}
R. J. Macfarlane, B. Lee, M. R. Jones, N. Harris, G. C. Schatz, and C. A. Mirkin,
 Science \textbf{334}, 204 (2011). 

\bibitem{Znag}
C. Zhang, R. J. Macfarlane, K. L. Young, C. H. J. Choi, L. Hao, E. Auyeung,
G. Liu, X. Zhou, and C. A. Mirkin, 
Nat. Mater. \textbf{12}, 741 (2013).

\bibitem{Isogai-jcg}
T. Isogai, A. Piednoir, E. Akada, Y. Akahosi, R. Tero, S. Harada, 
T. Ujihara, and M. Tagawa,
J. Cryst. Growth \textbf{401}, 494 (2014). 

\bibitem{Isogai-jjap}
 T. Isogai, E. Akada, S. Nakada, N. Yoshida, R. Tero, S. Harada, 
T. Ujihara, and M. Tagawa, Jpn. J. Appl. Phys. \textbf{55}, 03DF11 (2016). 




\end{thebibliography}
\end{document}